\documentclass[a4paper,preprint]{aastex}
\usepackage{graphicx}

\newcommand{\smfr}[2]{{\textstyle\frac{#1}{#2}}}

\newcommand{\wsq}{{\bf w}^2}
\newcommand{\Qd}[1]{Q^{[3]}_{#1}}
\newcommand{\Fr}[1]{F^{\mbox{\scriptsize #1}}}

\begin{document}

\title{Post Newtonian SPH} 

\author{
  S. Ayal\altaffilmark{1}, 
  T. Piran\altaffilmark{1},
  R. Oechslin\altaffilmark{2},
  M. B. Davies\altaffilmark{3}, and
  S. Rosswog\altaffilmark{4}
}
\altaffiltext{1}{Racah Institute for Physics, The Hebrew University, Jerusalem,
  Israel, 91904}
\altaffiltext{2}{Department f\"ur Physik und Astronomie, Universit\"at Basel,
  Switzerland}
\altaffiltext{3}{Department of Physics \& Astronomy, University of Leicester, UK}
\altaffiltext{4}{Center for Parallel Computing (ZPR / ZAIK), Universit\"at zu K\"oln, Germany}

\begin{abstract}
  We introduce an adaptation of the well known Tree+SPH numerical
  scheme to Post Newtonian (PN) hydrodynamics and gravity.  Our code
  solves the (0+1+2.5)PN equations. These equations include Newtonian
  hydrodynamics and gravity (0PN), the first order relativistic
  corrections to those (1PN) and the lowest order gravitational
  radiation terms (2.5PN). We test various aspects of our code using
  analytically solvable test problems. We then proceed to study the
  1PN effects on binary neutron star coalescence by comparing
  calculations with and without the 1PN terms. We find that the effect
  of the 1PN terms is rather small. The largest effect arises with a
  stiff equation of state for which the maximum rest mass density
  increases. This could induce black hole formation. The gravitational
  wave luminosity is also affected.
\end{abstract}

\keywords{gravitation,hydrodynamic,relativity,stars: neutron}

\section{Introduction} 

The development of numerical methods for the solutions of full 3D
general relativistic hydrodynamic problems is still at a preliminary
stage \citep{NGR}. In the meanwhile, various approximate approaches to
this problem have been attempted. Of these some include an
approximation to the metric in the form of the conformal flatness
condition (CFC) \citep{wilsmath} or by using a Post-Newtonian (PN)
formulation of the equations \citep{oohara,1997rggr.conf..309O}.
Recently it has been suggested that at least in some cases these two
approximations are of the same order of accuracy \citep{notWM}.  In
this work we set out to modify the popular Tree-SPH
\citep{1990ApJ...348..647B,1989ApJS...70..419H} numerical scheme
formulated for Newtonian self gravitating hydrodynamic systems to work
in the PN approximation.  We need to adapt both parts of the Tree-SPH
scheme - the tree code and the SPH. Our goal is to study the
coalescence of binary neutron stars (BNS), an intrinsically 3D
hydro+gravity problem, and to begin the exploration of the general
relativistic effects which are important in this process.

Smooth Particle Hydrodynamics (SPH) is a Lagrangian, grid-less
particle method for solving the hydrodynamic equations. It is this
lack of a grid which makes SPH especially appealing for the efficient
solution of complex 3D problems. SPH has been used in hydrodynamic
simulations including gravity, magnetic fields and in special
relativistic problems \citep{kmz,astro-ph/9904070}.
\citet{1993ApJ...404..678L} have also implemented a fully general
relativistic SPH with a fixed Kerr background. For a review on SPH
techniques see \citet{mon_rev} and \citet{benz_rev}. The Barnes-Hut
Tree algorithm \citep{bh} is a $O(N\log N)$ method for calculating
gravitational forces between $N$ particles. It has been combined with
SPH in order to get a powerful and efficient particle based
gravity+hydro solver.

We combine the efficiency of the SPH approach with a PN formalism
suggested by \citet{BDS} (BDS). The PN approximation to gravity
involves expanding the relativistic equations with a small parameter
$O(v^2/c^2)$ where $v$ is a typical velocity in the system (thermal
velocity, orbital velocity etc...). 1PN means we neglect all terms
proportional to $v^4/c^4$ and higher. The 1PN approximation includes
in the first order, many relativistic effects. The main shortcoming of
the 1PN approximation is that it misses all effects which have to do
with gravitational radiation.  These effects appear only at the 2.5PN
($v^5/c^5$) order. The BDS formalism includes the Newtonian physics
(0PN), the leading order PN effects (1PN) and the leading order
gravitational radiation effects (2.5PN) in a self consistent way.  It
recasts the PN equations in a form similar to the Newtonian equations
thus facilitating the adaptation of the Tree-SPH algorithm to solve
this problem.

In this work we introduce the code and present code tests. We also use
the code to simulate binary neutron star (BNS) coalescence in the
(0+1+2.5)PN approximation and to compare these results to Newtonian
simulations.  In section \ref{sec:num} we introduce the numerical
method.  In section \ref{sec:tests} we examine the results of various
code tests.  We present the results of the BNS coalescence simulation
in section \ref{sec:BNS} and we conclude in section \ref{sec:conc}.

\section{Numerical method\label{sec:num}}
The BDS formalism recasts
the (0+1+2.5)PN gravitation and hydrodynamic equations in a form
resembling the Newtonian (0PN) equations. This enables the solution of
these equations using methods adapted from Newtonian gravity (such as
the Tree+SPH method we use here). The formalism reduces all the
relativistic non-local equations to compact supported Poisson
equations. The PN order of the various terms in the equations of
motion can be read off their coefficients - 0PN terms have no
coefficients, 1PN have coefficients proportional to $1/c^2$ and 2.5PN
terms have coefficients proportional to $1/c^5$. This enables us to
``turn off'' various powers of the PN approximation by setting to zero
the corresponding coefficients. We use this option later when making a
(0+2.5)PN calculation.

The independent matter variables used are the following set: $\rho_*$
the coordinate rest mass density, $\varepsilon_*$ the coordinate specific
internal energy and $\bf{w}$ the specific linear momentum. In fully
relativistic terms these are defined as:
\begin{eqnarray}
  \rho_*  &=& \sqrt{g}u^0\rho,\\
  \varepsilon_* &=& \varepsilon(\rho_*),\\
  w_i &=& \left(c^2 + \varepsilon + p/\rho\right)\frac{u_i}{c},
\end{eqnarray}
where $\rho$ is the rest mass density, $\varepsilon(\rho)$ the
specific energy, $p(\varepsilon,\rho)$ the pressure and $u^\mu$ the
four-velocity (Greek indices run from 0 to 4, Latin indices from 1 to
3). The corresponding BDS variables are the above quantities
neglecting all terms except 0PN, 1PN and 2.5PN. Using these variables
the formalism yields an evolution system which consists of 9 Poisson
equations and 8 hyperbolic equations (see Appendix~\ref{sec:eqns}).

In the 0PN approximation ($c\rightarrow\infty$) $w_i=v^i$, $\rho_*$ is
the mass density, the only needed potential is $U_*$, and we recover
the known Newtonian equations of motion. From Eq.~(\ref{eq:Q}) we have
$\Qd{ij}=d^3Q_{ij}/dt^3+O(1/c^2)$, where $Q_{ij}$ is the quadropole
moment, enabling us to compute the gravitational radiation luminosity
of the system to lowest order as
\begin{equation}
  \label{eq:lum}
  L_{\rm GW} = \frac{1}{5}\frac{G}{c^5}\Qd{ij}\Qd{ij}.
\end{equation}

The BDS equations are self-consistent as long as the system is only
mildly relativistic. This can be characterized by noting that at least
the parameters $\alpha/c^2$, $\beta/c^2$ and $\delta/c^2$ (Eq.
\ref{eq:alpha}, \ref{eq:beta} and \ref{eq:delta} respectively) must be
small. This means that contrary to 0PN systems which are always
self-consistent but can be physically wrong the 1PN system will cease
to be self consistent at the time when its results are too far from
the general relativistic results. This inconsistency can be understood
as follows: given two functions expanded in a small parameter
$\eta \ll 1$
\begin{eqnarray}
  f &=& f_0 + f_1\eta + \ldots,\\
  g &=& g_0 + g_1\eta + \ldots,
\end{eqnarray}
the product of these functions is
\begin{equation}
  fg = f_0g_0 + (f_0g_1 + f_1g_0)\eta + (f_0g_2 + 2f_1g_1 +
  f_2g_0)\eta^2+\ldots.
\end{equation}
If we choose to truncate the functions at the first order (setting
$f_i=0$ for $i>1$) and indeed $\eta\ll 1$ then we can also neglect the
$\eta^2$ terms in the product, but if $\eta\approx 1$, we will also
need to take into consideration the $2f_1g_1\eta^2$ term in the
product if we wish to be self-consistent (the approximation would
break down in any case giving wrong results). Similarly in order to
make the 1PN system self consistent we would need to add some 2PN
terms to the equations complicating them even more. This is not the
case if we wish to truncate the functions at the zero order ($f_i=0$
for $i>0$). Then we can consistently truncate the product also at the
zero order. This explains the self-consistency of the Newtonian (0PN)
approximation.

For the solution of the Poisson equations we used a Barnes-Hut tree
\citep{bh}. The Barnes-Hut tree is a $O(N\log N)$ method commonly used
for $N$-body gravitational force calculations. The principle behind
the Barnes-Hut tree is that at a given position, the gravitational
force from a cluster of distant particles can be approximated by using
only global properties of the cluster - the monopole, dipole and
quadrupole moments of the mass distribution. The Barnes-Hut tree
provides a simple way to calculate these moments and to cluster the
particles. It is remarkable that the same formalism can be used with
minor modification to solve any compact supported elliptic equation.
We adapt this method to solve a general compact supported elliptic
equation (see Appendix~\ref{sec:ap1}) and we use it here to solve
numerically equations (\ref{eq:Us})-(\ref{eq:R}). In all our runs we
set the tree accuracy parameter $\theta$ to 0.5. Gravitational
softening was naturally incorporated by assigning particles with their
respective SPH mass density in the gravitational calculations.

The evolution equations were solved using SPH, a Lagrangian particle
based scheme for solving hydrodynamic problems. The use of SPH was
facilitated by the similarity of the equations in the BDS formalism to
the Newtonian equations (indeed this is the whole point in the BDS
formalism). The particle mass used in Newtonian SPH was replaced by
the conserved mass $m_* = \int d^3x\, \rho_*$ in our scheme. When
using SPH Eq.\ \ref{eq:cont} is automatically satisfied and only
equations\ \ref{eq:ener}-\ref{eq:moment} have to be solved. The use of
SPH requires adding some artificial viscosity in order to resolve
shocks. We use the standard artificial viscosity
\citep[e.g.][]{mon_rev,benz_rev} consisting of a term analogous to
bulk viscosity and a Von Neuman-Richtmyer artificial viscosity term.

We used a symmetrical form of the SPH equations which guarantees the
exact conservation of the ``momentum'' $\int d^3x\, \rho_*w_i$ in the
hydrodynamics section of the code. The momentum isn't exactly
conserved in the gravity section of the code because the tree
algorithm introduces a small asymmetry into the forces. This problem
arises also in Newtonian SPH calculations as well and it leads to
spurious accelerations of the center of mass of the system. These
acceleration are small for a typical Tree parameters (of the order of
$10^{-6}$ compared to other accelerations in the system) and can be
corrected by simply subtracting the center of mass acceleration from
the particle accelerations at each time step.

For time integration we used a second order Runge-Kutta integrator
with adaptive step size control. The step size was determined so that
none of the integrated variables will change by more than a
predetermined amount, set to 0.5\% in all runs. In addition we check
that the Courant condition is always satisfied. We did not implement
single particle time steps. The simulations were run on standard
Pentium II 450MHZ workstations and took about 255 hours of CPU time
($\sim 11$ days) each.
\section{Code Tests\label{sec:tests}}
Numerous studies have been conducted on the capabilities and
limitations of the SPH formalism
\citep{1989ApJS...70..419H,1993A&A...272..430D,lombardi99:}.  We have
put our code through a suite of tests to insure that our
implementation is correct and that it reproduces previously obtained
results using available codes.  Specifically we have compared the
Newtonian (0PN) version to the Newtonian SPH code used in
\cite{mbd_piran}.  This comparison ensures the validity of the
Newtonian part of our code.  We devised other simple tests for the
different PN aspects of the code - (0+1)PN hydrodynamics,
gravitational radiation damping (2.5PN) and (0+1)PN hydrodynamics and
gravity. These all have an analytical or easily obtainable result and
take a reasonable time to run.

\begin{figure}
  \begin{center}
    a\includegraphics{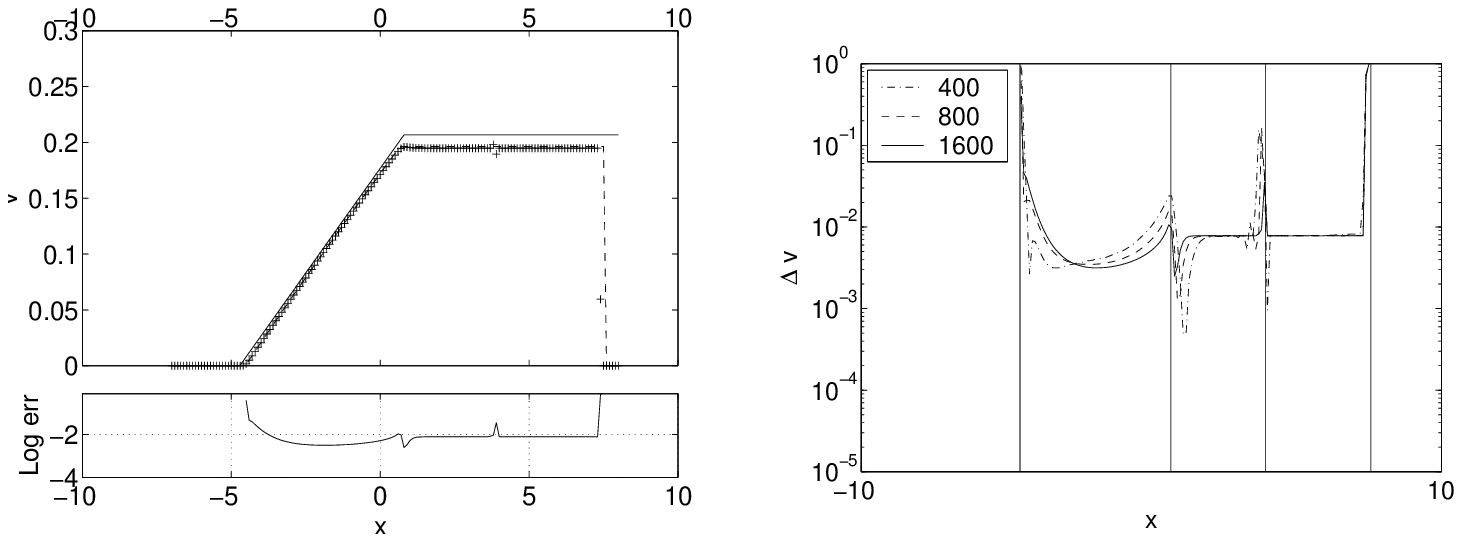}\\
    b\includegraphics{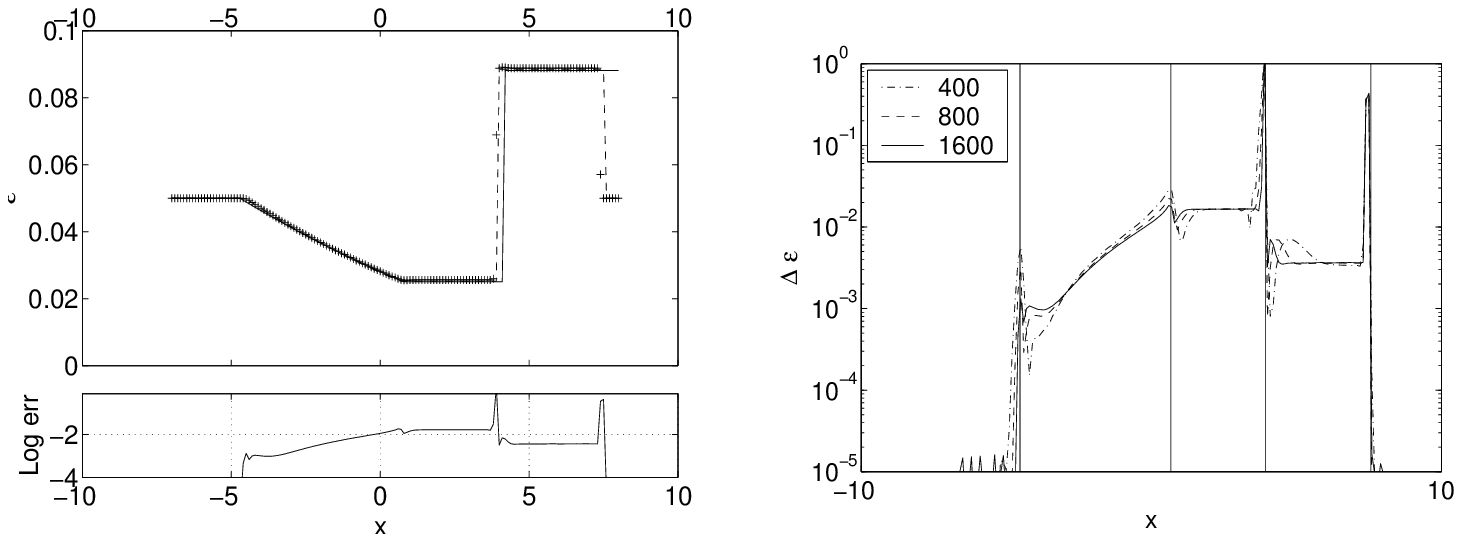}\\
    c\includegraphics{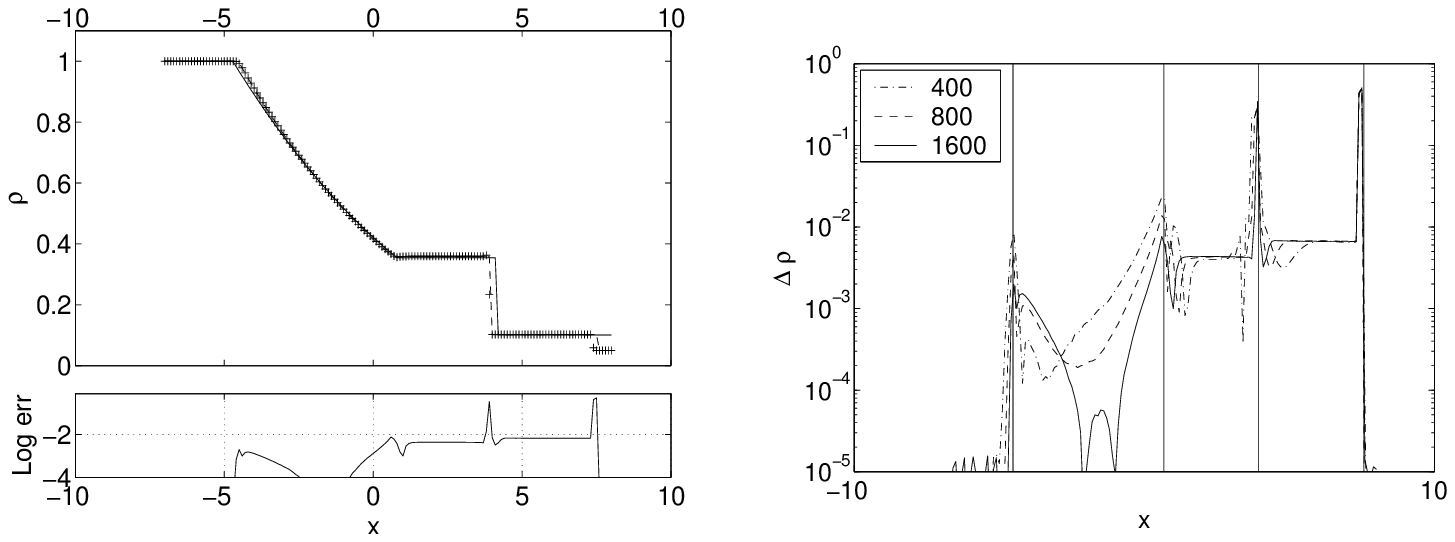}
    \caption{
      \label{fig:shock_tube}
      Shock tube results for a $\gamma=5/3$ gas. (a) The velocity. (b)
      The internal energy density $\varepsilon$. (c) The rest mass
      density $\rho$. Initial conditions were $\rho_{*l} = 1$,
      $\rho_{*r} = 0.05$, and $\varepsilon_* = 0.05$ using 800
      particles. On the left, each graph shows the Newtonian
      analytical results in a solid line, the relativistic analytical
      results in a dashed line and the (0+1)PN numerical points as
      crosses. Under each graph is the error in the (0+1)PN result as
      compared to the relativistic result. On the right the each graph
      shows the error relative to the analytic relativistic solution.
      The vertical lines mark the left edge of the rarefraction, the
      right edge of the rarefraction, the contact discontinuity and
      the shock going from left to right.  The error is calculated for
      solutions with 400, 800 and 1600 particles.  Except around the
      discontinuities the error converges to $\approx 10^{-2}$, which
      is the expected error resulting from neglecting 2PN and higher
      terms ($\varepsilon^2 \approx \cdot 10^{-2}$).}
  \end{center}
\end{figure}

(0+1)PN hydrodynamics (without gravity) were tested using the 1D shock
tube problem. We have compared the results to both the Newtonian and
the relativistic solutions
\citep{taub48,1973ApJ...181..903M,1984ApJ...277..296H}. These tests
were conducted with a 1D version of our code which differs from the 3D
version used elsewhere in this paper only by the normalization of the
SPH smoothing kernel. The results of the tests are presented in Figure
\ref{fig:shock_tube}. These tests demonstrate that our PN code gives
better results than a Newtonian code for mildly relativistic problems,
with shock velocities of $\sim 0.2c$.  These conditions are at the
upper limit of the validity of the (0+1)PN approximation as
$\varepsilon\approx 0.1c^2$. We compare (0+1)PN calculation to the
analytic Newtonian solution and the relativistic analytical solution.
The error in the (0+1)PN velocity is smaller by an order of magnitude
than the Newtonian error. This leads to a more accurate estimate of
the shock's position. The errors for the energy density and mass
density are larger, but still better or comparable to the Newtonian
error. We conclude that even in these extreme conditions for the
(0+1)PN hydrodynamic approximation, it fares at least as well as
Newtonian hydrodynamics, and is much better at estimating the velocity
of the fluid. In all the quantities, the relative error of the (0+1)PN
result as compared to the relativistic analytical solution is of the
order of 1\% except for single particles which reside at
discontinuities.  In Figure (\ref{fig:shock_tube}) we also show the
convergence of the results.  We show the relative error compared to
the analytic relativistic results for different particle numbers.
Except at the discontinuities the error relative to the (0+1)PN result
converges to $\approx 10^{-2}$.  The size of the zone around the
discontinuities with a large error decreases when the number of
particles increases, indicating, again, that the discontinuities
affect only a fixed small ($\approx 10$), number of particles. The
convergence of the error to $\approx 10^{-2}$ is consistent with the
order of magnitude of the largest 2PN term - $\varepsilon^2/c^4$ which
is the expected error of the (0+1)PN approximation. We point out that
3D tests of shocks with the particles positioned randomly, which are
expected to model the conditions in a 3D run more realistically, give
much poorer resolution \citep{1991ApJ...377..559R}.

\begin{figure}
  \begin{center}
    \includegraphics[width=10cm]{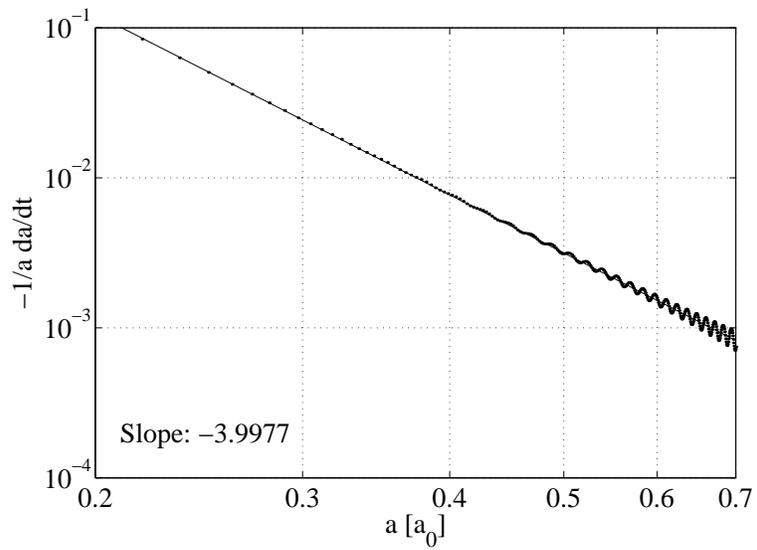}
    \caption{
      \label{fig:grdamp}
      Results of (0+2.5)PN runs with point masses which check the
      accuracy of gravitational radiation damping: a plot of
      $-\dot{a}/a$ vs. $a$. The initial points with $a>0.7a_0$ (where
      $a_0$ is the initial separation) were excluded to allow for
      numerical relaxation. The expected slope is -4.}
  \end{center}
\end{figure}

Gravitational radiation backreaction (the 2.5PN terms) was tested
using two point masses in orbit. In this run, all the 1PN terms were
discarded leaving only (0+2.5)PN terms. Since we used two point
masses, hydrodynamics didn't play any role in this test. The rate of
change of the orbit's radius was compared to the analytical result of
$\dot{a}/a\propto -a^{-4}$ (see e.g. \citet{bhwdns}, Chapter 16).  
Figure~\ref{fig:grdamp} depicts  the results of  this test. The two point
masses were set initially in a Keplerian orbit. Since this is only an
approximation to actual quasi-stationary orbits of (0+2.5)PN
gravitation, a brief period of relaxation preceded the orbital decay
after which the expected power law emerged.

The (0+1)PN gravity and hydrodynamics were tested using a static
polytropic star with a polytropic index $\gamma=5/3$. Initial
conditions were constructed using a (0+1)PN expansion of the
Oppenheimer--Volkoff (OV) equations for a relativistic spherical star.
We ran this test with two resolutions, 2151 and 5140 particles
respectively, for approximately 50 hydrodynamic time scales (which we
take to be $\sqrt{R^3/GM}$). for the initial conditions we use
regularly placed unequal mass particles. The stars were constructed by
placing the particles in a face-centred cubic (FCC) lattice, thus
maximizing the number of neighbors each particle has.  Using an
iterative procedure, each particle is than assigned a mass so that the
resulting density profile matches the OV density profile.

\begin{figure}[]
  \begin{center}
    a\includegraphics[width=12cm]{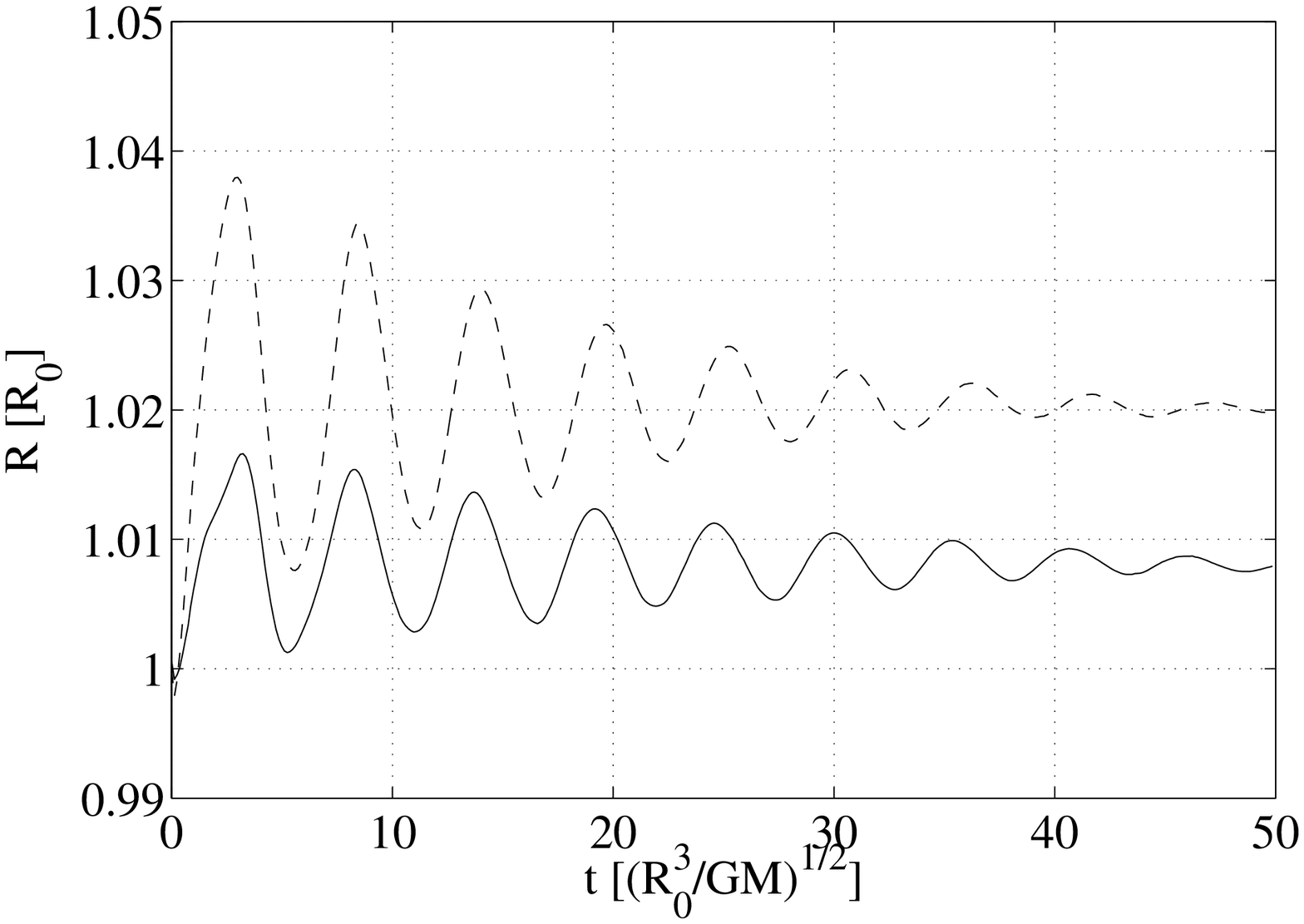}\\
    \vspace{0.5cm}
    b\includegraphics[width=12cm]{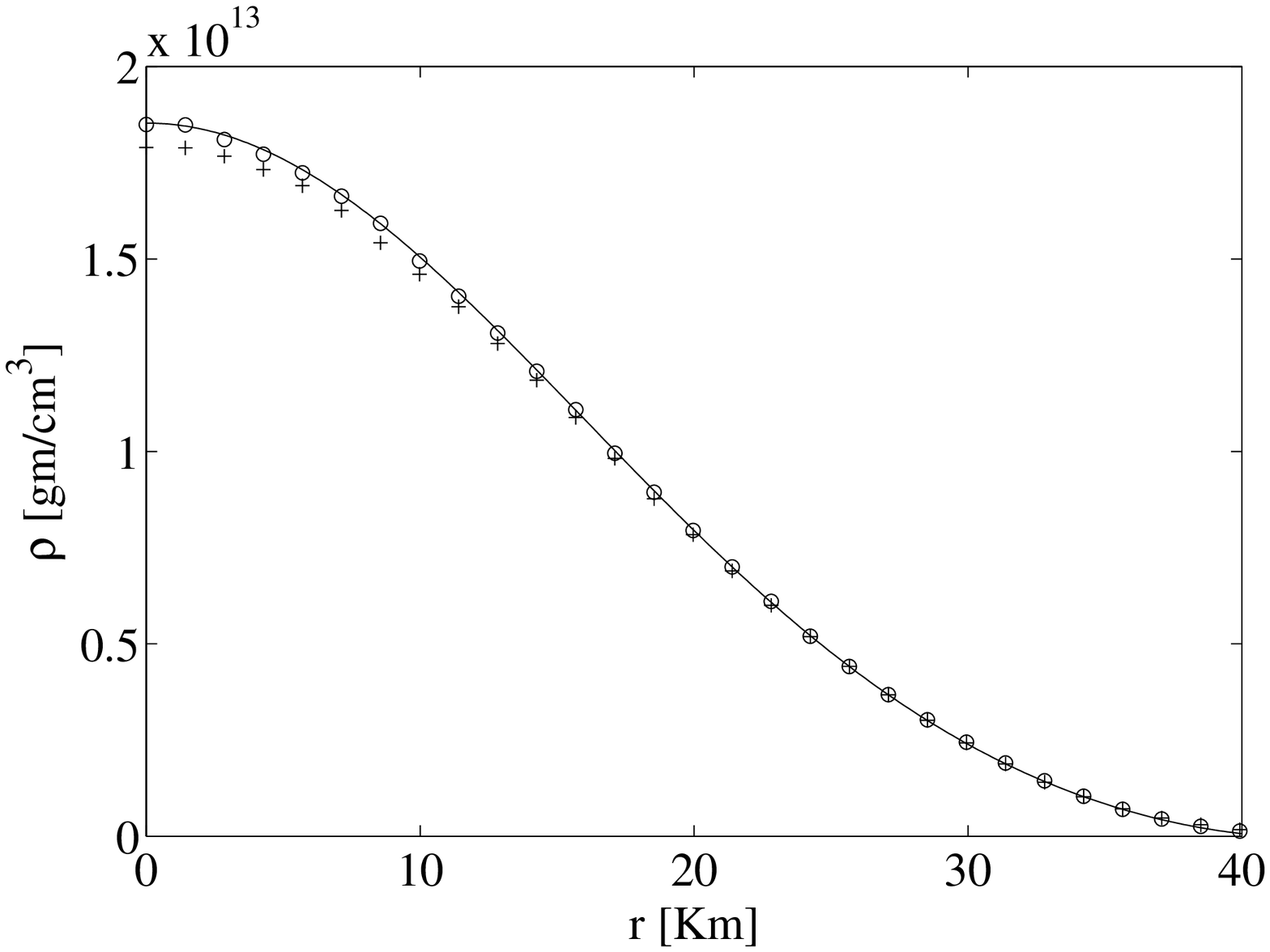}
    \caption{
      \label{fig:pn_poly}
      (0+1)PN $\gamma=5/3$ polytropic stars made out of 2151 and 5140
      particles. The simulation ran for about 50 hydrodynamic
      time scales. The parameters of the star were $R_0=34$~Km and
      $M=0.49~{\rm M_\odot}$. (a) The radius inclosing 95\% of the mass. The
      solid line and dashed line are for the 2151 and 5140 particle
      runs respectively. (b) Rest mass density profiles. The initial
      profile in a solid line and the final profiles of the 2151 and
      5140 particle runs in crosses and circles respectively.}
  \end{center}
\end{figure}

Figure~\ref{fig:pn_poly}a depicts the radius enclosing 95\% of the
mass.  This radius oscillates with a decreasing amplitude converging
to a constant value. There are less than 50 oscillation in the graphs
since the oscillation period for these stars is some factor of order
unity times $\sqrt{R^3/GM}$, which we take to be the hydrodynamical
timescale. The oscillations are caused by several factors in the SPH
algorithm. Every SPH particle adjusts it's size $h$ so that it will
have approximately 50 other particles as neighbors (within a sphere of
radius $2h$). In our initial conditions all the particles have the
same $h$, this causes the particles near the star's surface to have
only half the neighbors, and their $h$ is increased by the code. This
in turn lowers the density and takes the star out of equilibrium.
Another reason for the oscillations is that $\nabla P \ne 0$ on the
boundary of the star due to numerical errors. Finally in the initial
conditions the particles are positioned on a Cartesian grid which
clearly conflicts with the spherical symmetry of the OV density
profile assigned to them. The star begins to oscillate but the
oscillations are damped by the artificial viscosity present in the SPH
algorithm. The expected errors in all quantities (radius, central
density, etc...) from the SPH discretisation error $(\langle h\rangle
/R_0)^2$ are 1.5\% and 0.9\% for the 2151 and 5140 particle runs
respectively. As can be seen in Figure~\ref{fig:pn_poly}, the
oscillations decrease when we increase the resolution. The final rest
mass density, shown in Figure~\ref{fig:pn_poly}b, converges towards
the analytical curve as the resolution increases.

\section{PN BNS coalescence\label{sec:BNS}}
BNS's provide an excellent test-bed for gravitational and nuclear
astrophysics. A binary pair of neutron stars will loose angular
momentum and energy via gravitational wave emission as has been
observed \citep[see][and references therein]{taylor_nobel}. This
process will ultimately lead to a coalescence of the two neutron
stars.  The gravitational waves emitted from the coalescence are
expected to be observed by gravitational wave detectors coming on-line
in the next decade, such as LIGO \citep{1998nwap.conf..391B}, VIRGO
\citep{1997grgp.conf..163F}, GEO \citep{1996magr.meet.1352H} and TAMA
\citep{1999eama.confE..I3K}.

At the final stages of the coalescence the BNS system must be
described using detailed 3D modeling of gravitational and
hydrodynamic effects. This restricts the study of the last stages of
coalescence to numerical methods. Many groups have performed numerical
simulations using different approximations to this problem. Results
have been obtained using Newtonian dynamics by \citet{mbd_piran},
\citet{RS2}, \citet{swesty}, \citet{ruffert} and \cite{stephan}. Post
Newtonian (PN) results have also been obtained by \citet{oohara} and
\citet{1997rggr.conf..309O}. Recently an almost fully relativistic result
was obtained by \citet{wilsmath} who used the CFC approximation
(although see \citet{errWM} and \citet{notWM} for cautionary remarks
on the validity of this approximation).

Although sophisticated techniques were developed in order to obtain
equilibrium binary configurations both in the (0+1)PN
\citep{shibata1,shibata2} and general relativistic \citep{baumgarte}
cases, converting the resulting density field into SPH particles is
not a trivial task. Previous works using Newtonian SPH usually
manufacture equilibrium configurations by relaxing the system in the
co-rotating frame, in which the stars are stationary
\citep[e.g.][]{RS1}. This approach was unavailable to us because of
the complicated nature of the (0+1+2.5)PN equations. Instead we assign
two spherical stars with Keplerian velocity (as in
\citealt{shibata98:_stabil}), and after some oscillations the system
settles down into a stationary state.

We model the binary NS system by two equal polytropes with zero spins
relative to an inertial observer (For further discussion on the
initial spins and their implications see \citet{stephan} and
references therein). On top of this we made several other simplifying
assumptions in our calculation which are already discussed in
\cite{mbd_piran}, namely ignoring neutrino transport. The masses we
use for each star are less than $\rm M_\odot$ for radii of about 30
Km. Although these parameters are far from realistic for NS's, they
allow us to investigate the effect of general relativity on the
coalescence while still in the regime of applicability of the
(0+1+2.5)PN approximation for the whole duration of the run. We
compare the (0+1+2.5)PN results (hereafter denoted P) to (0+2.5)PN
runs (hereafter denoted N) with the same mass and initial separation
to highlight the 1PN effects. We make a total of 6 runs (3 pairs of
P-N runs). The parameters for these different runs are summarized in
Table \ref{tab:parms}.

\begin{table}[htbp]
  \begin{center}
    \begin{tabular}{|l|c|c|c|c|c|c|c|c|}\tableline
      Run & $\gamma$ & M [$\rm M_\odot$] & 
      \multicolumn{2}{|c|}{$R_*$ [Km]} & Particles &
      \multicolumn{2}{|c|}{$m_{\rm esc}$ [$\rm M_\odot$] (\# part.)}
      \\\cline{4-5}\cline{7-8}
         &          &      & P    & N    &      & P & N \\\tableline\tableline
      1  &  $5/3$   & 0.52 & 29.8 & 33.9 & 4996 & $<3.7\times 10^{-5}$ (0) & $<3.7\times 10^{-5}$ (0) \\\tableline
      2  &   2.6    & 0.50 & 34.7 & 36.9 & 20562& $2.6\times 10^{-4}$ (8)& $6.9\times 10^{-4}$ (20)\\\tableline
      3  &   2.6    & 0.92 & 30.5 & 35.6 & 18520& $<4.3\times 10^{-5}$ 
      (0) &  $4.5\times 10^{-4}$ (6)\\\tableline
    \end{tabular}
    \caption{
      Summary of results and parameters for the various runs. M is the
      conserved mass per star, $R_*$ the radius encompassing 95\% of
      the conserved mass of each star in the initial configuration,
      $m_{\rm esc}$ is the mass of unbound particles in the final
      configuration. The upper limits on $m_{\rm esc}$ are the masses
      of the lightest particles in the run.
      }
    \label{tab:parms}
  \end{center}
\end{table}

\begin{figure}
  \begin{center}
    \includegraphics[width=10cm]{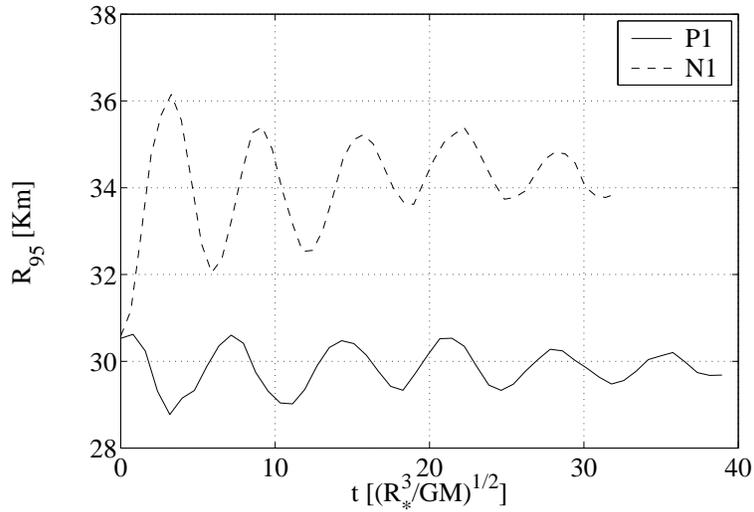}
    \caption{
      The radii of the stars as a function of time for the beginning
      of the P1 and N1 runs. The initial polytropes are static in the
      (0+1)PN approximation. The initial expansion in the N run
      reflects in the larger size of the N static polytrope.  Based on
      this graph we take the radius of the P1 run to be 29.8Km and the
      radius of the N1 run to be 33.9Km. For the P1 run this is within
      3\% of the initial radius and for the N1 run this is 10\% more
      than the initial radius. Note that in this graph the units are
      different than in all following graphs}
    \label{fig:rad_c}
  \end{center}
\end{figure}

\begin{figure}
  \begin{center}
    a\includegraphics[width=12cm]{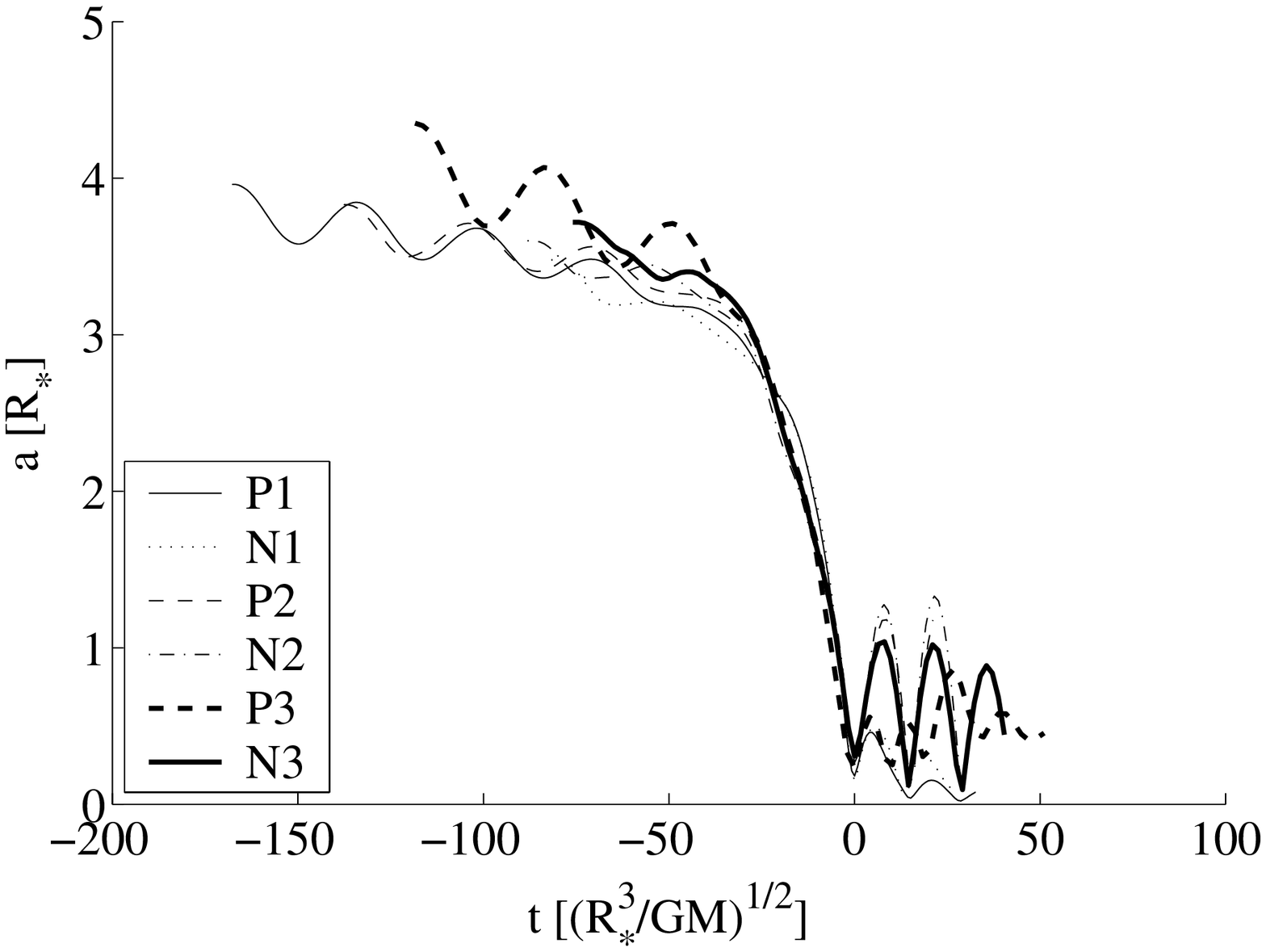}\\
    \vspace{0.5cm}
    b\includegraphics[width=12cm]{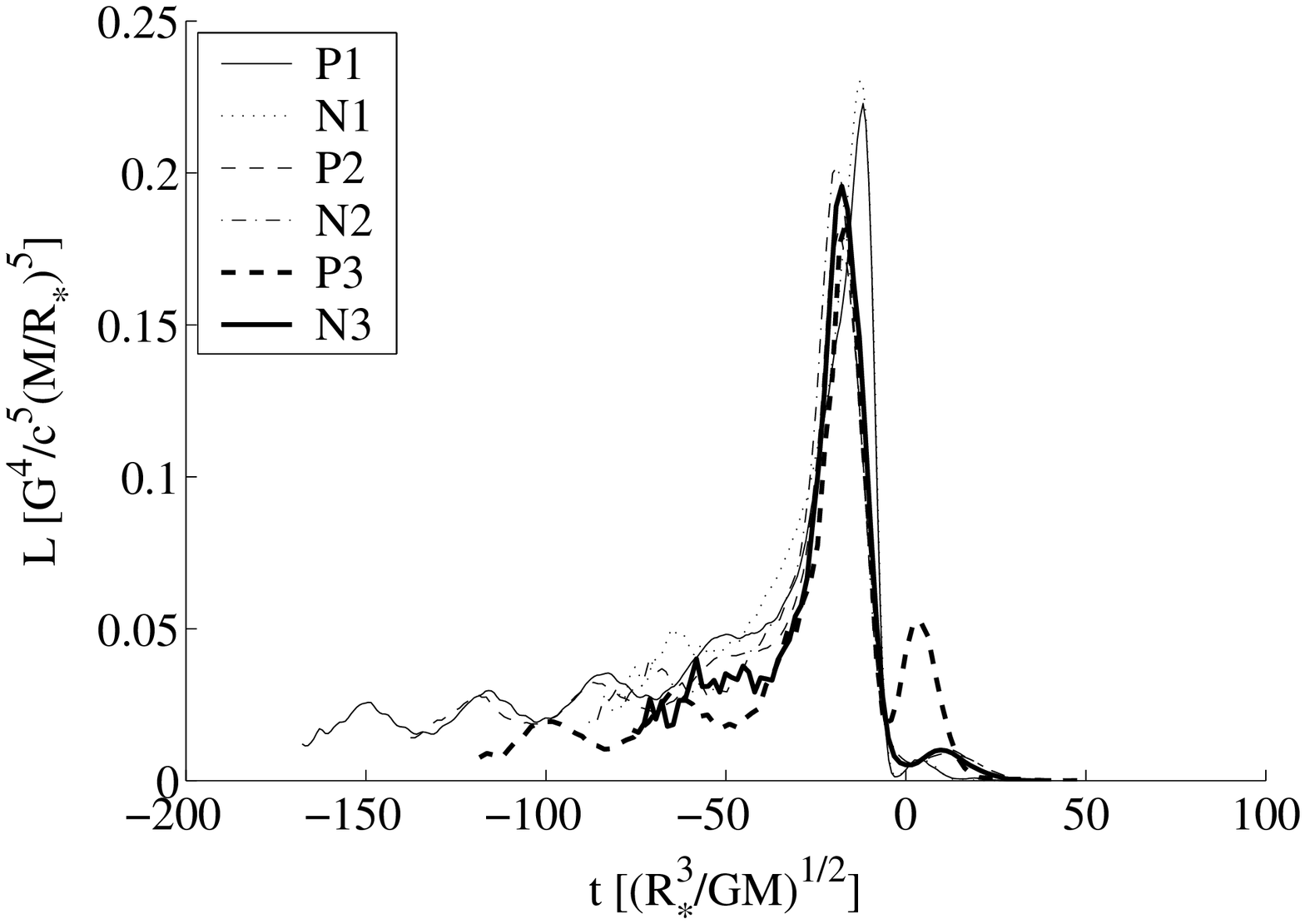}
    \caption{
      (a) The separation between the centers of mass of the stars (in
      units of initial radius) as a function of time for all the runs.
      (b) The gravitational wave luminosity of the system for all the
      runs. 
      }
    \label{fig:seplum}
  \end{center}
\end{figure}

When comparing a P run to a N run with the same mass, we must take
into account that a polytrope that is static in (0+1)PN gravitation is
not static in 0PN, Newtonian, gravitation. For the same mass, the
Newtonian polytrope has a larger radius. This causes the initial
polytrope to grow at the beginning of the N runs as demonstrated in
Figure ~\ref{fig:rad_c}. Note that unlike Figure~\ref{fig:pn_poly}a
where we show the radii of isolated starts, here we show the radii of
the stars in the first timesteps of the coalescence. In order to
differentiate between 1PN effects and the effects of these differences
in initial conditions, we try to present all results scaled by the
appropriate mass and initial radius.  For instance, the dynamical
orbital instability sets in at $a/R_*\approx 3$ \citep{RS2} where $a$
is the binary separation. For a N run this is at a larger separation
than for a P run with similar mass. This means that the P runs will
have to evolve longer with gravitational radiation damping as the only
mechanism for decreasing the separation.  This behavior can be seen in
Figure~\ref{fig:seplum}a.

The scaling we use during the rest of the discussion is the following:
distance is measured in units of $R_*$, luminosity is measured in
units of $G^4/c^5(M/R_*)^5$, rest mass density ($\rho_*$) is measured
in units of $M/R_*^3$ and energy in units of the rest mass energy of
the star $Mc^2$. Time is measured in hydrodynamic time scales
$(R_*^3/GM)^{1/2}$ and is shifted so that for each run the minimal
separation occurs at $t=0$. $R_*$ is the initial radius of the stars.
$M=\int d^3x \rho_*$ is the {\em conserved} mass for each run. We
emphasize that in the P runs the conserved mass is not the same as the
rest mass.

\begin{figure}
  \begin{center}
    \includegraphics[width=12cm]{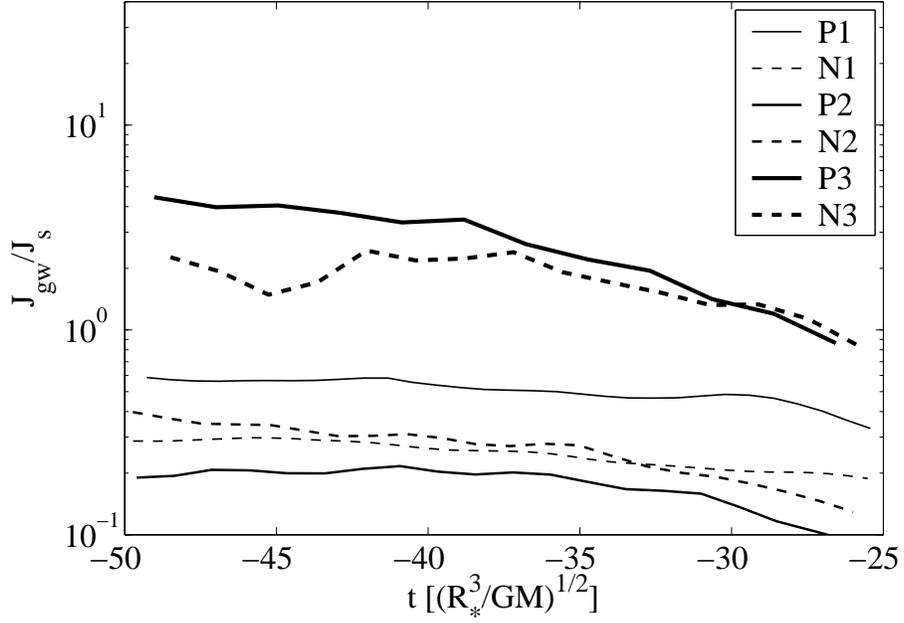}
    \caption{
      The ratio $J_{\rm gw}/J_{\rm s}$ of angular momentum in the
      gravitational radiation to angular momentum in stellar spins.
      The ration is shown up to the time of the dynamical orbital
      instability. In all runs but N3 and P3 spin-up of the stars is
      the dominant process for causing the merger.}
    \label{fig:J}
  \end{center}
\end{figure}

We define the center of mass of each star to be the center of {\em
  conserved} mass of the particles belonging to the star at the
initial time step. Since we use a particle based scheme we can follow
particles throughout the simulation and use this definition even after
the stars have touched.  In Figure~\ref{fig:seplum}a we show the
separation of the centers of mass of the two stars as a function of
time. As can be seen, the runs start in a slightly elliptical orbit
which slowly decays. Two distinct processes cause this decay. The
first is the conversion of orbital angular momentum into stellar spins
$J_{\rm s}$ via gravitational torques. This process is also present in
Newtonian gravity. The second process is the emission of gravitational
radiation, carrying with it angular momentum $J_{\rm gw}$. This is a
2.5PN process having no Newtonian parallel. The dynamical orbit
instability sets in for both the P and N runs as the separation
reaches $3R_*$, ($t\approx -25$). This causes a rapid plunge, and a
merger in about one orbital period. In Figure~\ref{fig:J} we show
the ratio $J_{\rm gw}/J_{\rm s}$ up to the dynamical instability. In
the runs N3 and P3 this ratio is close to unity near $t=-25$. In all
other runs it is apparent that the spin-up of the stars plays a more
important role than gravitational radiation in causing the merger. The
stars touch at $t \approx -15$ (when $a/R_*\approx 2$).  The minimal
separation at $t=0$ is achieved when the cores, which hold most of the
mass, have merged.  At $t>0$ the centers of mass ``bounce''. The
bounce is more pronounced for the P2, N2 and N3 runs because of the
stiffer EoS used in these runs.  In the P3 run however, 1PN gravity is
strong enough to counteract this stiffness and the bounce is
comparable to that of the P1 and N1 runs.

In Figure~\ref{fig:seplum}b we depict the gravitational radiation
luminosity (Eq. \ref{eq:lum}) of the merger.  The characteristics of
the luminosity peak are similar in all the runs once we allow for the
different initial radii of the stars. The P3 run however exhibits a
pronounced second peak at $t\approx 5$ at about the same luminosity of
the system at the last orbits before the merger. This second peak is
absent in all other runs. This distinct feature is a result of the
strong 1PN gravity and stiff EoS of the P3 run.

\begin{figure}
  \begin{center}
    a\includegraphics[width=9cm]{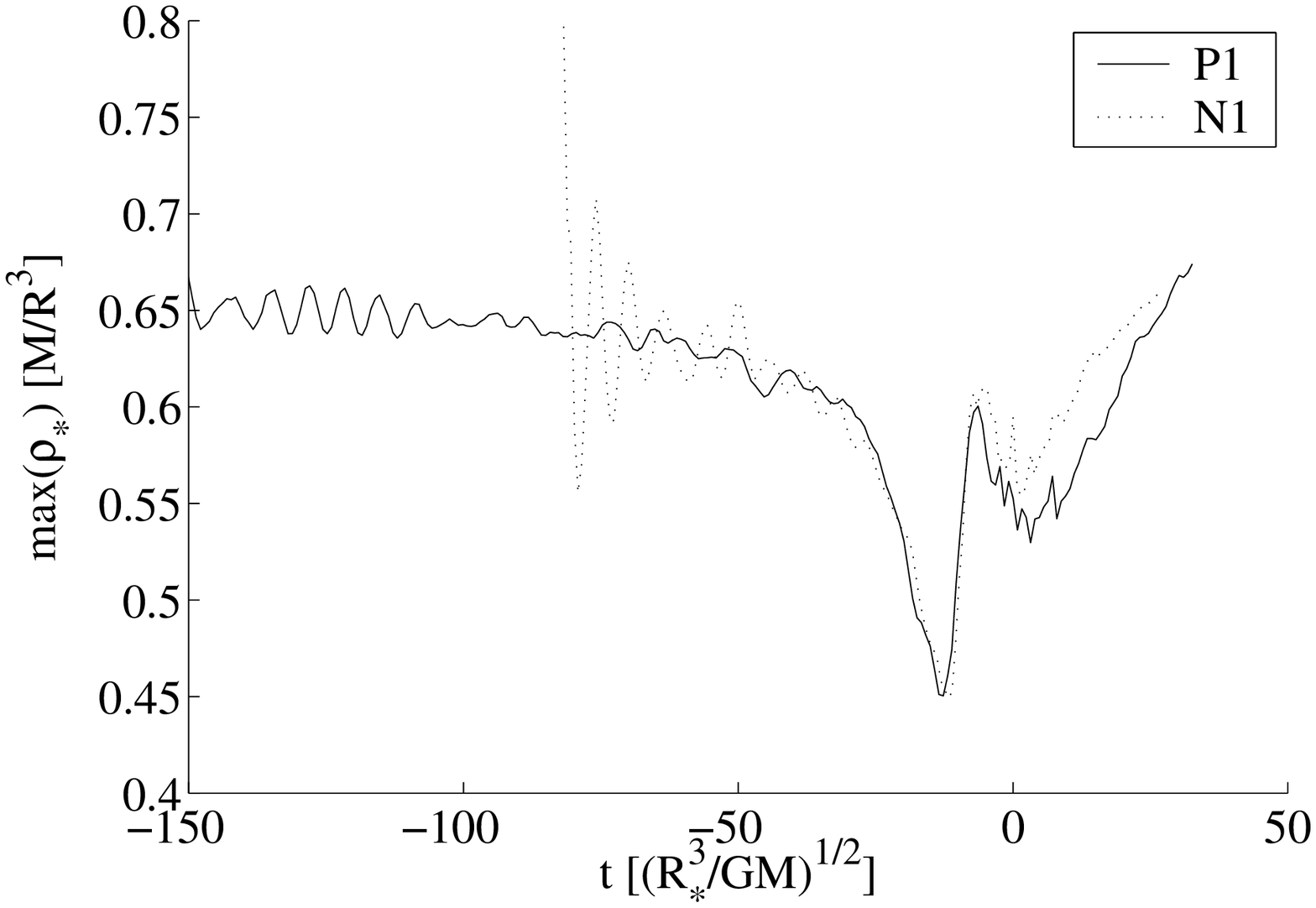}\\
    \hspace{0.5cm}
    b\includegraphics[width=9cm]{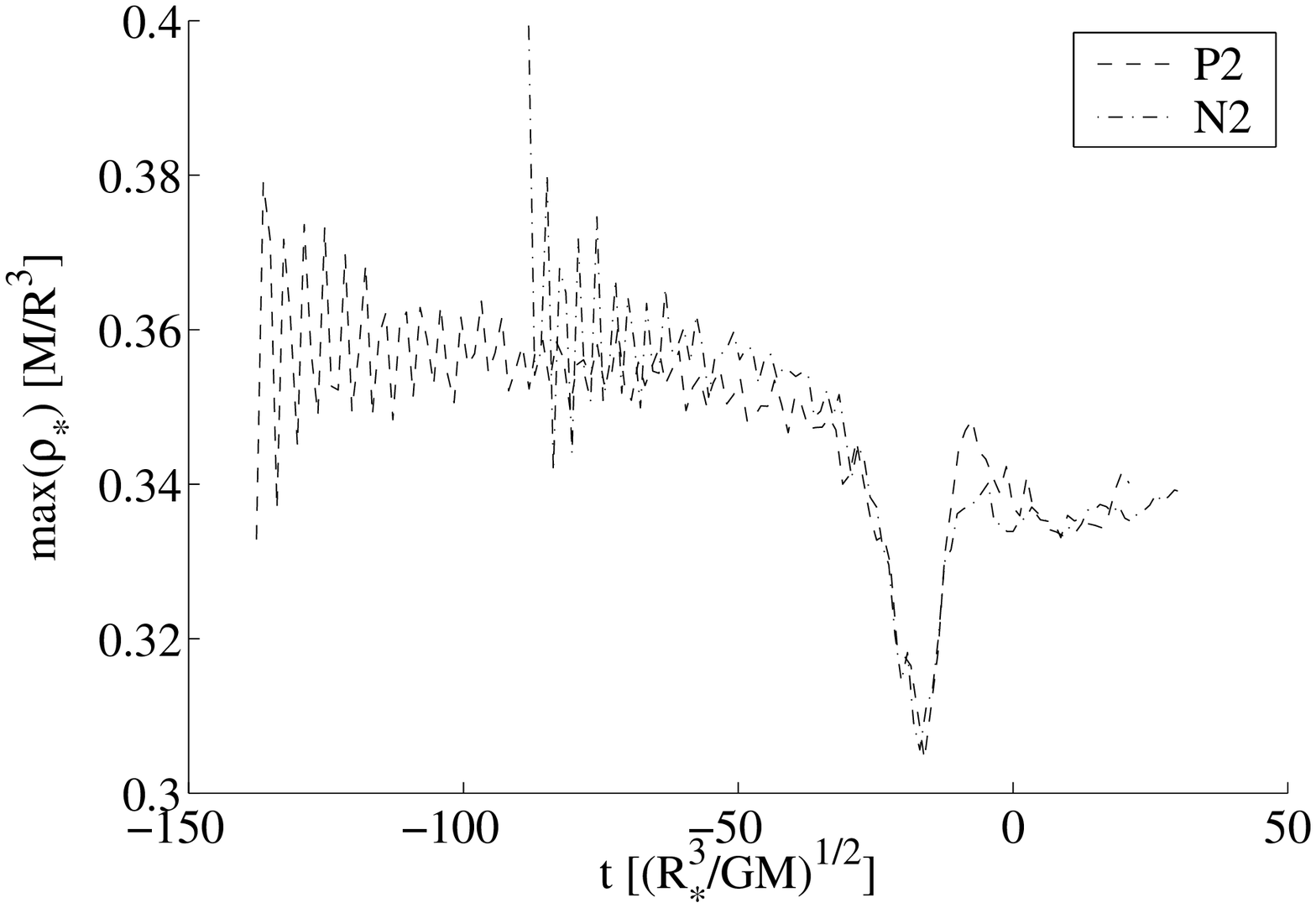}\\
    \hspace{0.5cm}
    c\includegraphics[width=9cm]{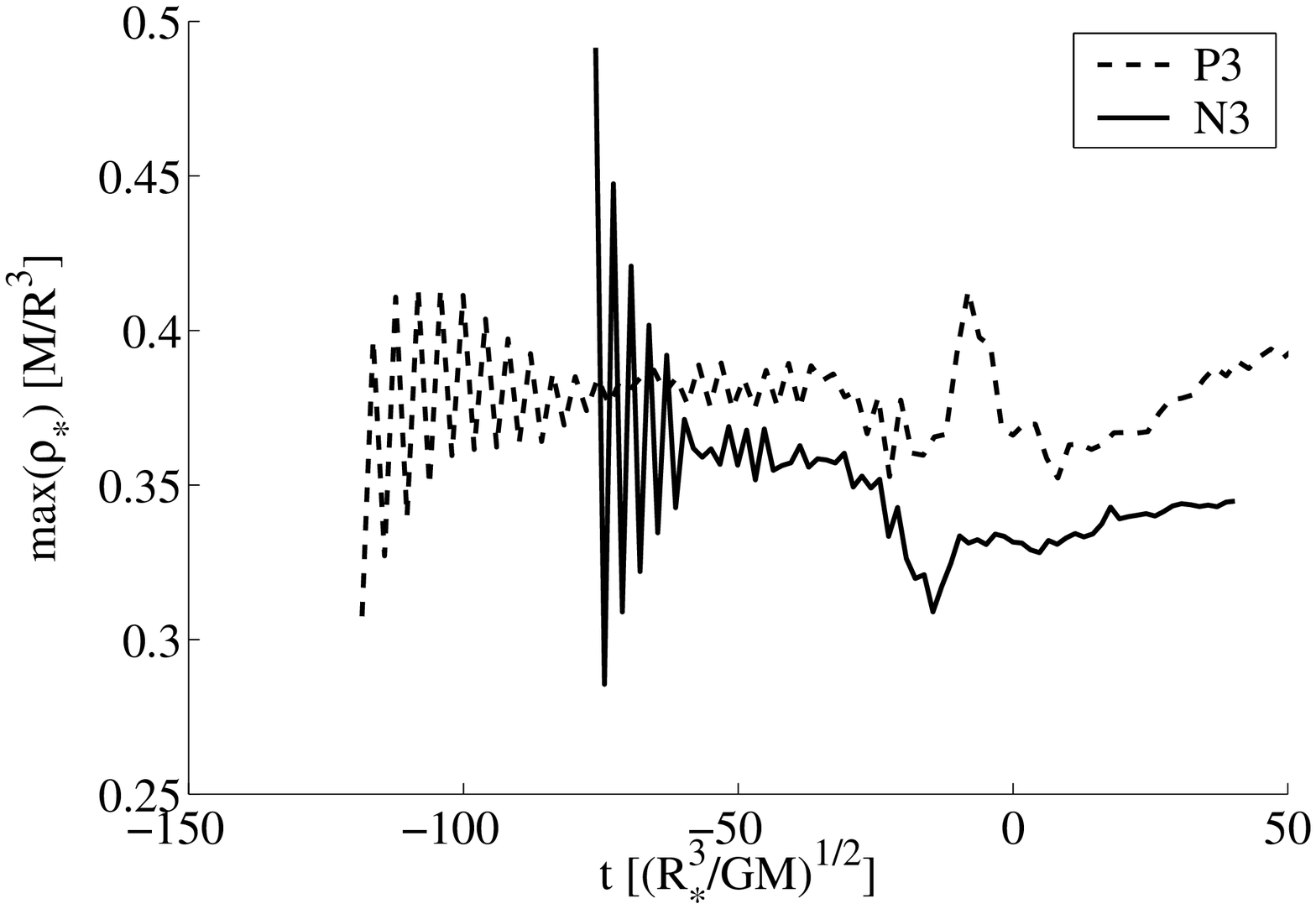}    
    \caption{
      Maximum coordinate rest mass density $\rho_*$ in the P and N
      runs. (a) the P1 and N1 runs. (b) The P2 and N2 runs. (c) The P3
      and N3 runs. The maximum drops during the merger and climbs back
      up afterwards. Notice the distinct rise at $t\approx -5$ in the
      P3 run }
    \label{fig:mrs}
  \end{center}
\end{figure}

\begin{figure}
  \begin{center}
    a\includegraphics[width=5cm]{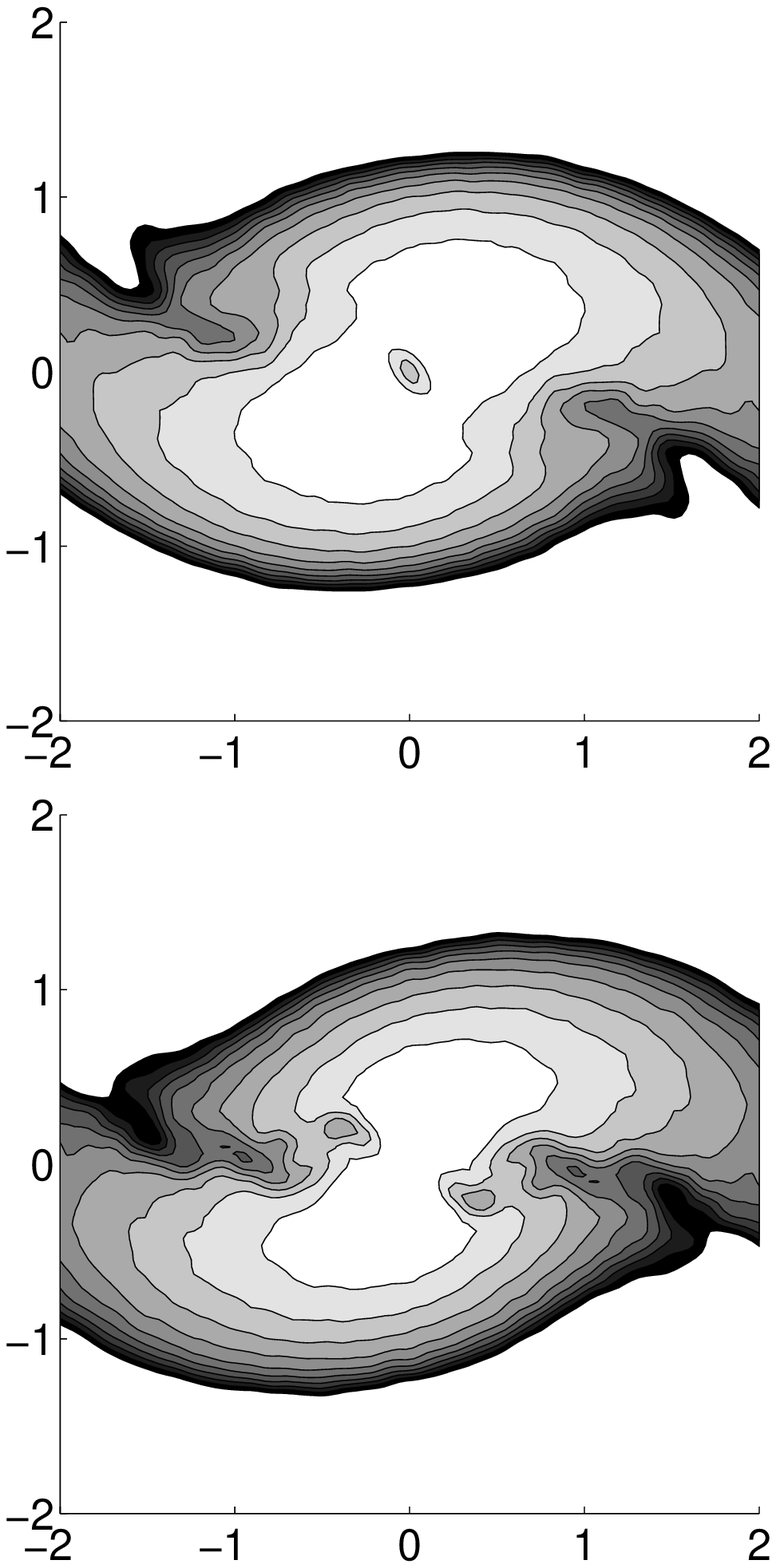}    
    b\includegraphics[width=5cm]{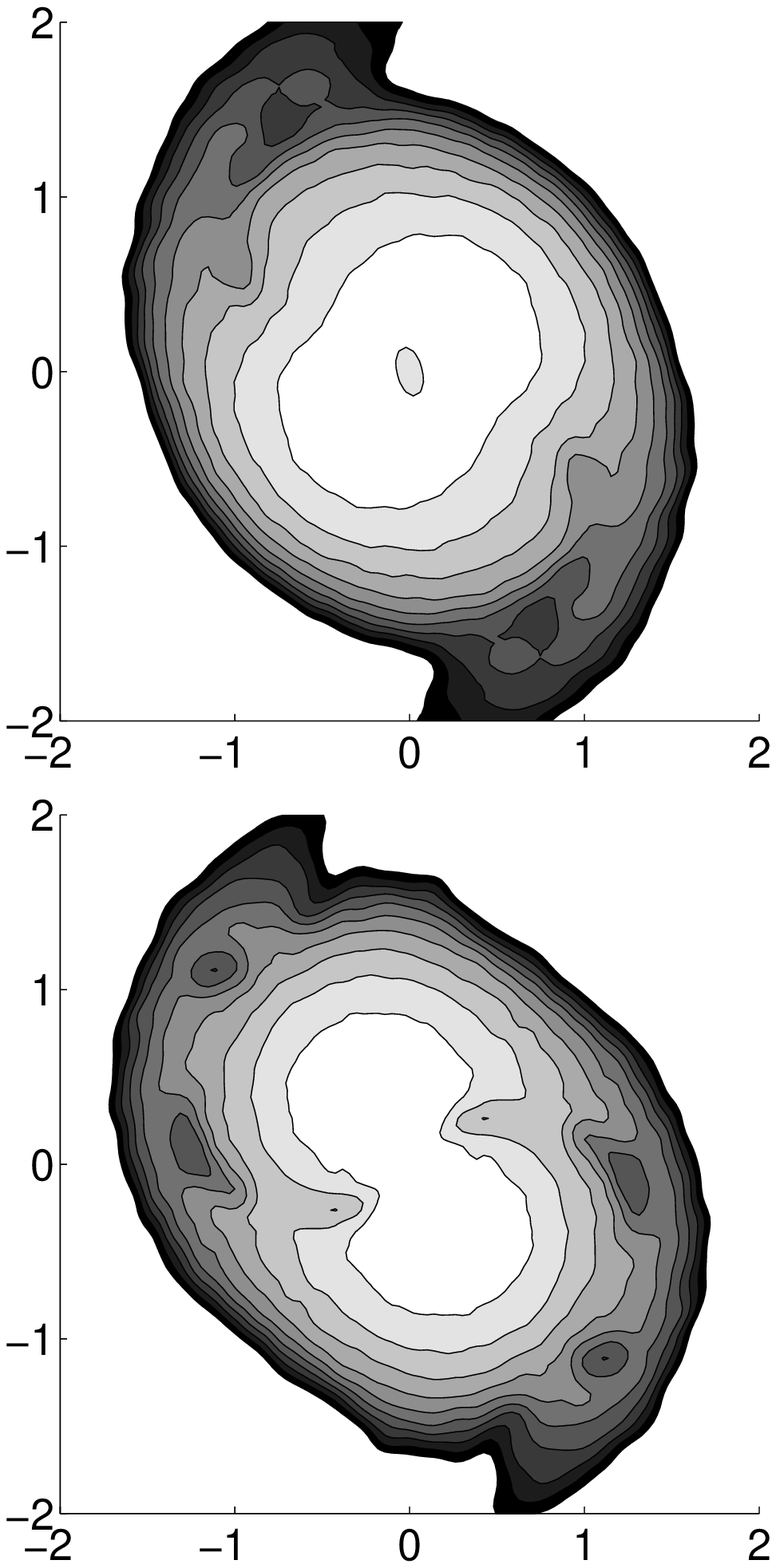}    
    c\includegraphics[width=5cm]{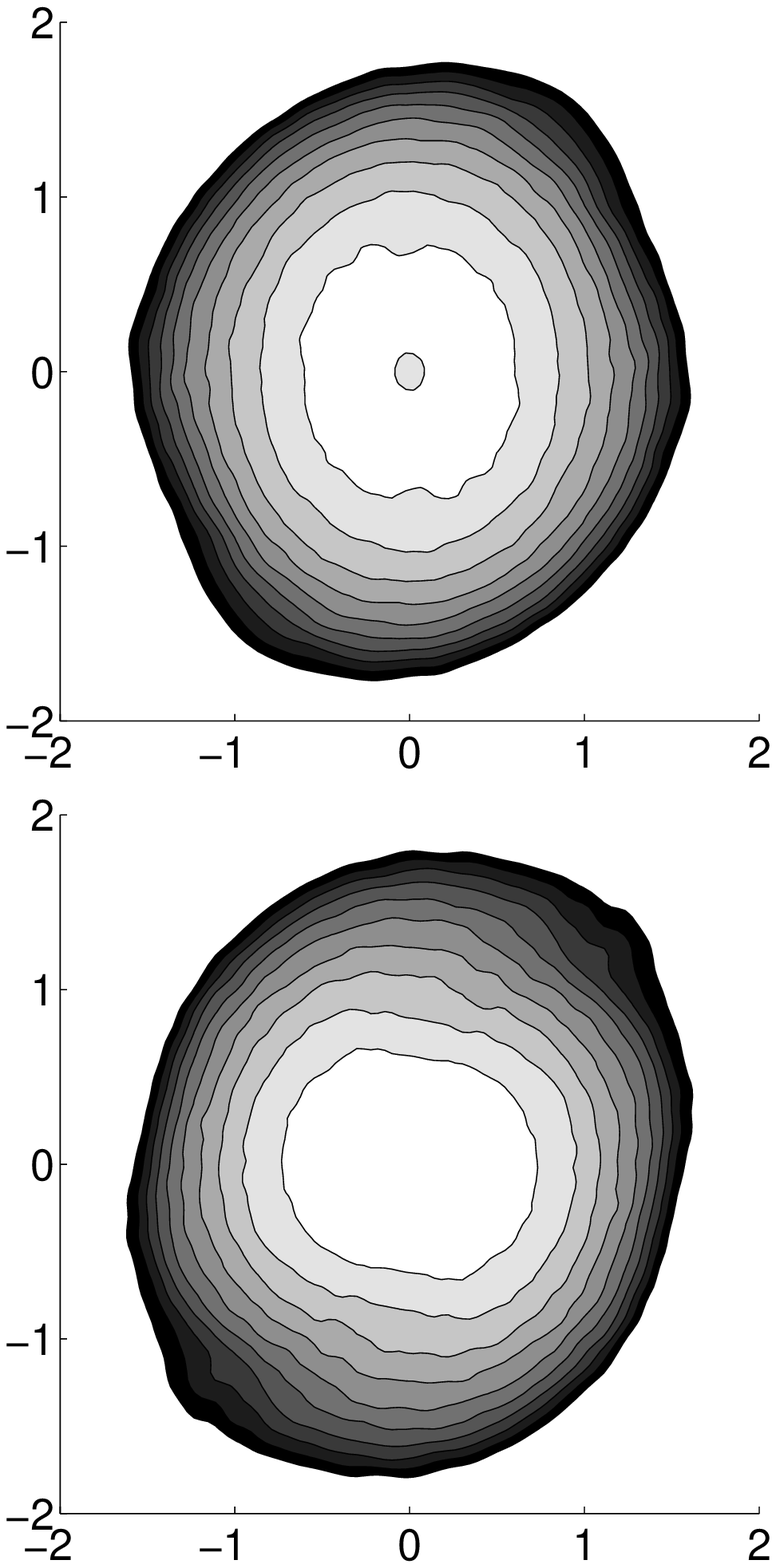}
    \caption{
      Rest mass density $\rho_*$ of the cores of the N3 (top) and P3
      (bottom) run. The contours are logarithmic with a spacing of
      $10^{0.1}$. The length scale is in $R_*$. The minimal contour is
      at 10\% of the maximum rest mass density so the spiral arms are
      not visible in this plot.  The cores is shown at times (a)
      $t\approx -10$. (b) $t\approx 2$. (c) $t\approx 20$.  }
    \label{fig:p3cont}
  \end{center}
\end{figure}

In order to explain this distinct feature of the P3 run we show in
Figure~\ref{fig:mrs} the maximum rest mass density $\rho_*$ for all
the runs. The P1 and N1 runs exhibit a larger relative density because
of their softer EoS. In all the runs we see a dip in the maximum
density at the time of the rapid infall caused by the orbital
instability. This corresponds to the stars shedding each other's mass
as they move closer together. This stage ends at $t\approx -15$, when
the stars touch, and is followed by a fast rise up to $t\approx -7$.
The difference between the N2 and N3 runs and the N1 run can be
attributed to the softer EoS of the latter while the difference
between them and their respective P runs is due to the stronger
gravitational attraction in the P runs. The maximum density in the N2
and N3 runs does not have the peak at $t\approx -7$ which is evident
in all other runs, most distinctly in the P3 run where it rises to
about 10\% more than it's final value. It is in this run where the
stiff EoS and the strong 1PN gravity combine to induce a large
compression of the cores which delays their final merger into a single
axisymmetric central object. This delay turns the merger into a two
part process and produces the second peak in the luminosity at
$t\approx 3$. This can be seen in Figure~\ref{fig:p3cont} where we
compare the cores of the P3 and N3 runs at $t\approx -10,2,20$.
$t\approx -10$, is just after the first peak in the luminosity. We see
the in the N3 run the cores have almost merged.  At $t\approx 2$ the
cores in the N3 run have merged and formed an almost axisymmetric
object which emits very little gravitational radiation, on the other
hand the P3 cores are still almost separate and certainly far from
axial symmetry. At $t\approx 20$ the cores of the P3 run have already
merged completely and emit little gravitational radiation as well.

In this context it is interesting to note \cite{1997A&A...321..991R}
where there is a comparison of the gravitational wave luminosity
produced by different numerical schemes.  SPH is found to inhibit the
second peak in the gravitational wave luminosity possibly because of
the numerical viscosity. Using other numerical schemes the second peak
in the luminosity is about one half the height of the first peak. This
raises the possibility that 1PN terms, may in reality have an even
more prominent effect on the luminosity.

\begin{figure}
  \begin{center}
    a\includegraphics[width=12cm]{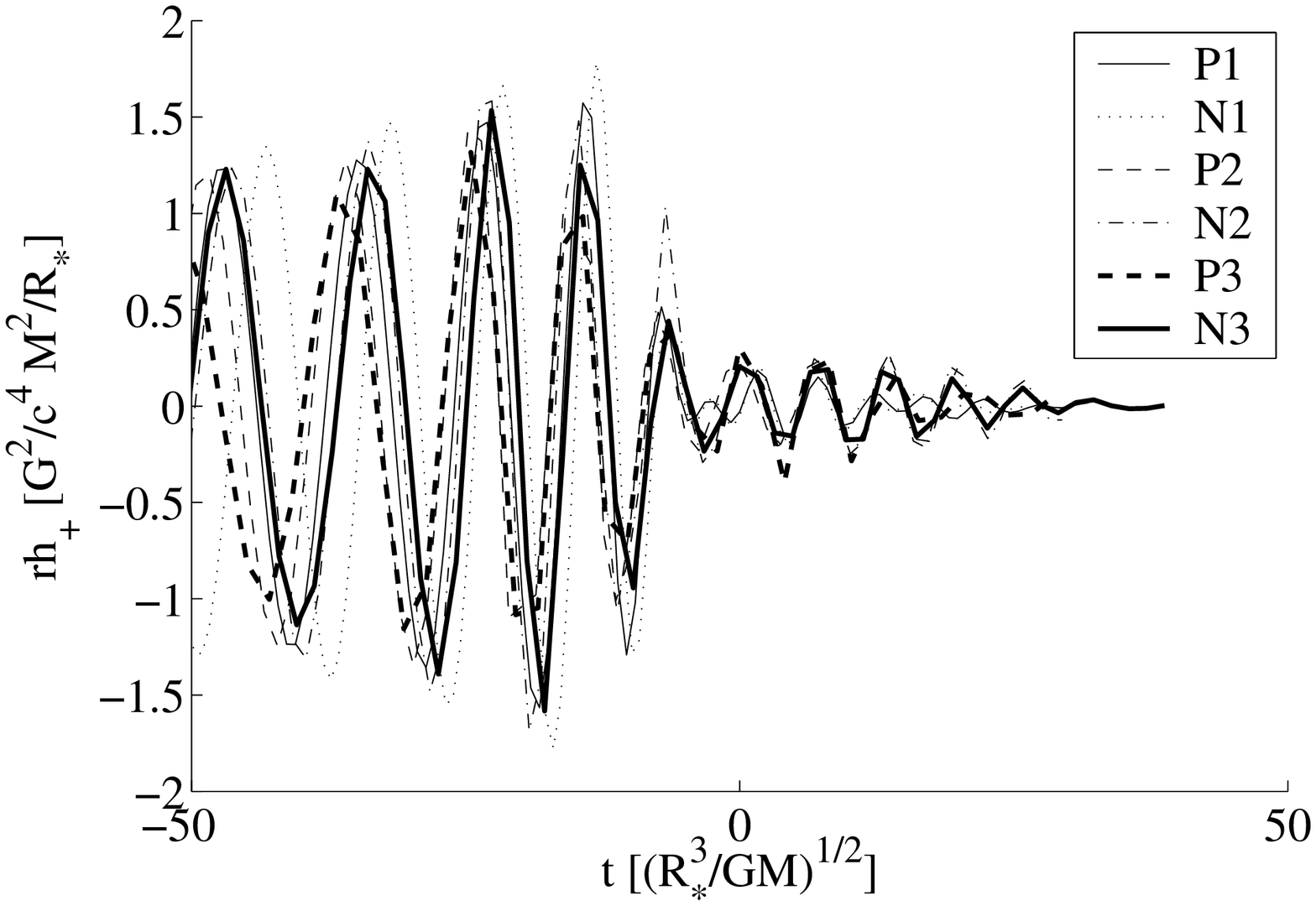}\\
    \vspace{0.5cm}
    b\includegraphics[width=12cm]{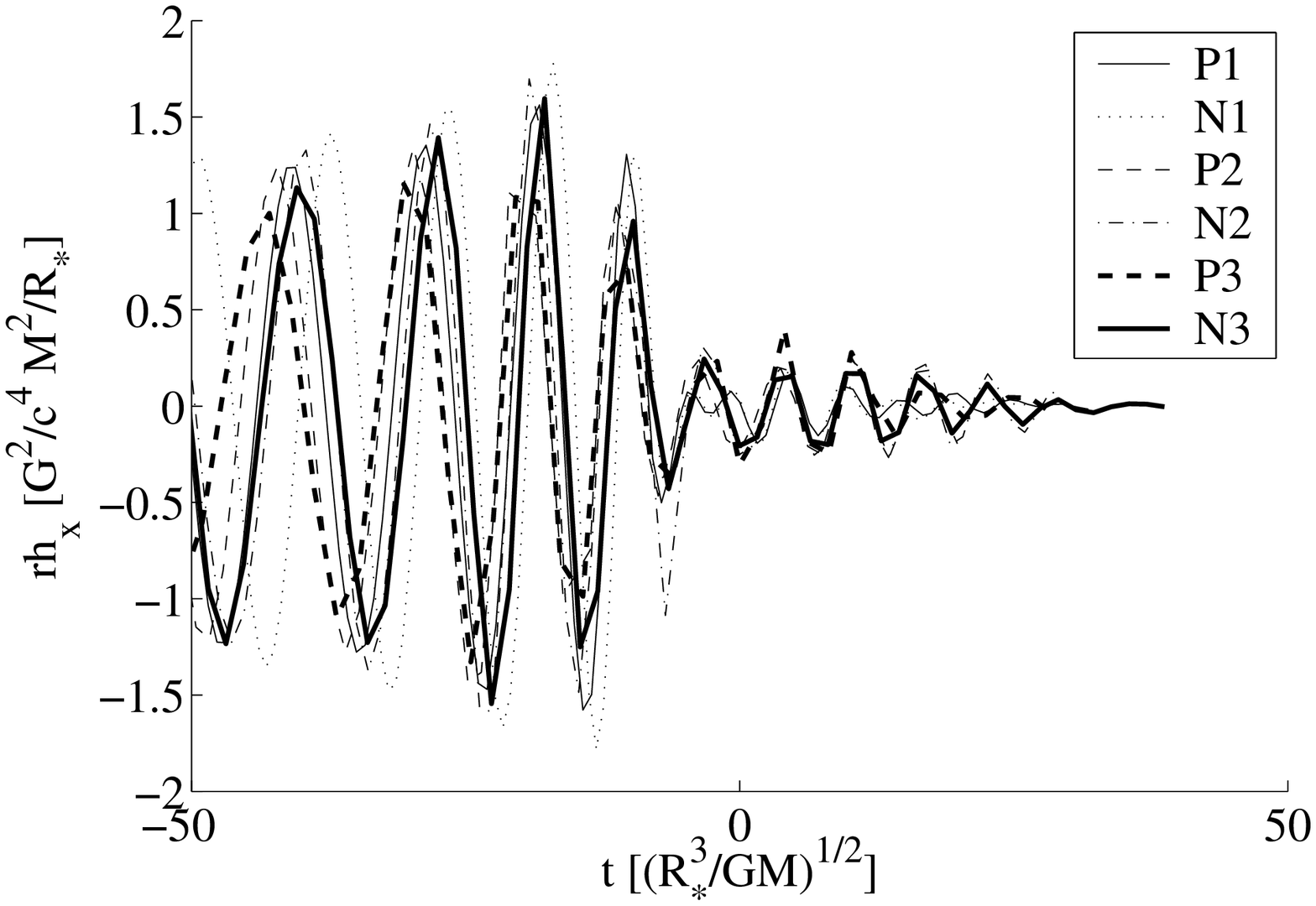}
    \caption{
      The gravitational radiation waveforms for an observer situated
      along the rotation axis (z axis) in geometrical units as a
      function of time. (a) The $+$ polarization. (b) The $\times$
      polarization }
    \label{fig:hr}
  \end{center}
\end{figure}

The actual gravitational waveforms emitted by the systems are shown in
Figure~\ref{fig:hr}. Here we see a difference in the period of the
waveforms between the P and N runs corresponding to a different
orbital period before the actual merger. During and after the merger
there is no qualitative difference between the waveforms.

\begin{figure}
  \begin{center}
    \includegraphics[width=12cm]{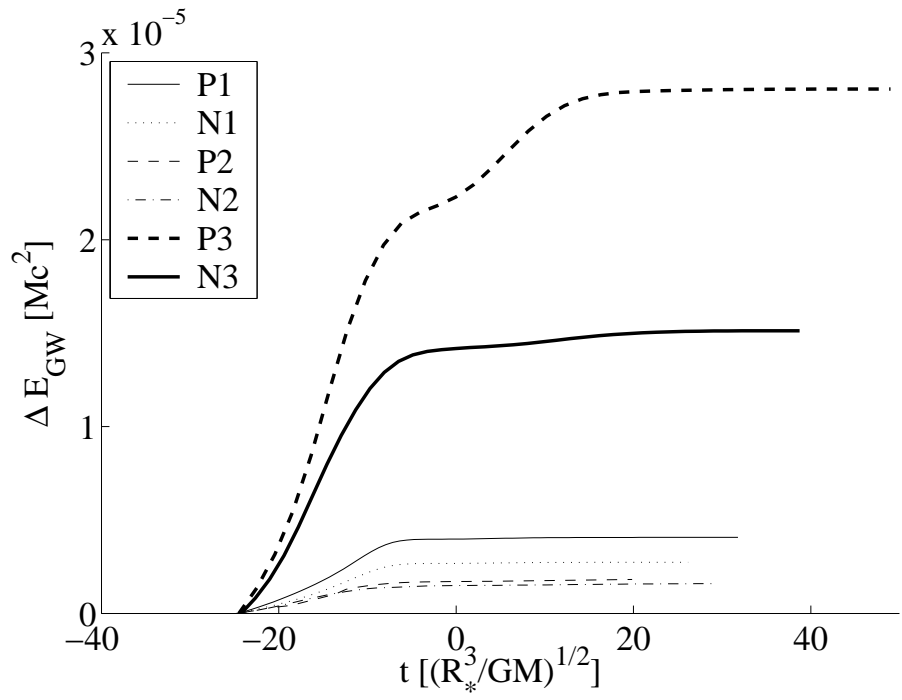}
    \caption{
      The energy emitted in gravitational waves for the P and N runs.
      The calculation was started at $t=-25$ when the runs had about the
      same relative separation.}
    \label{fig:cumlum}
  \end{center}
\end{figure}

In Figure~\ref{fig:cumlum} we compare the total energy emitted by
gravitational waves. We start the comparison at $t=-25$ when all runs
have roughly the same relative separation, before the stars touch.
When comparing the similar mass runs P1, N1, P2, and N2 we see that a
softer EoS implies more energy emitted in gravitational waves, while
the more massive runs P3 and N3 emit an order of magnitude more energy
as compared to run P2 and N2 which have a similar EoS. The P3 run
emits almost twice the energy of the N3 run. We note that in this
figure the axis scaling is different from Figure~\ref{fig:seplum}b
e.g. run P3 and N3 have almost similar $L$ in Figure~
\ref{fig:seplum}b, but since it is scaled by $R_*^5$ the actual P3
luminosity is a factor of $\approx 2$ higher than the N3 luminosity

\begin{figure}
  \begin{center}
    \includegraphics{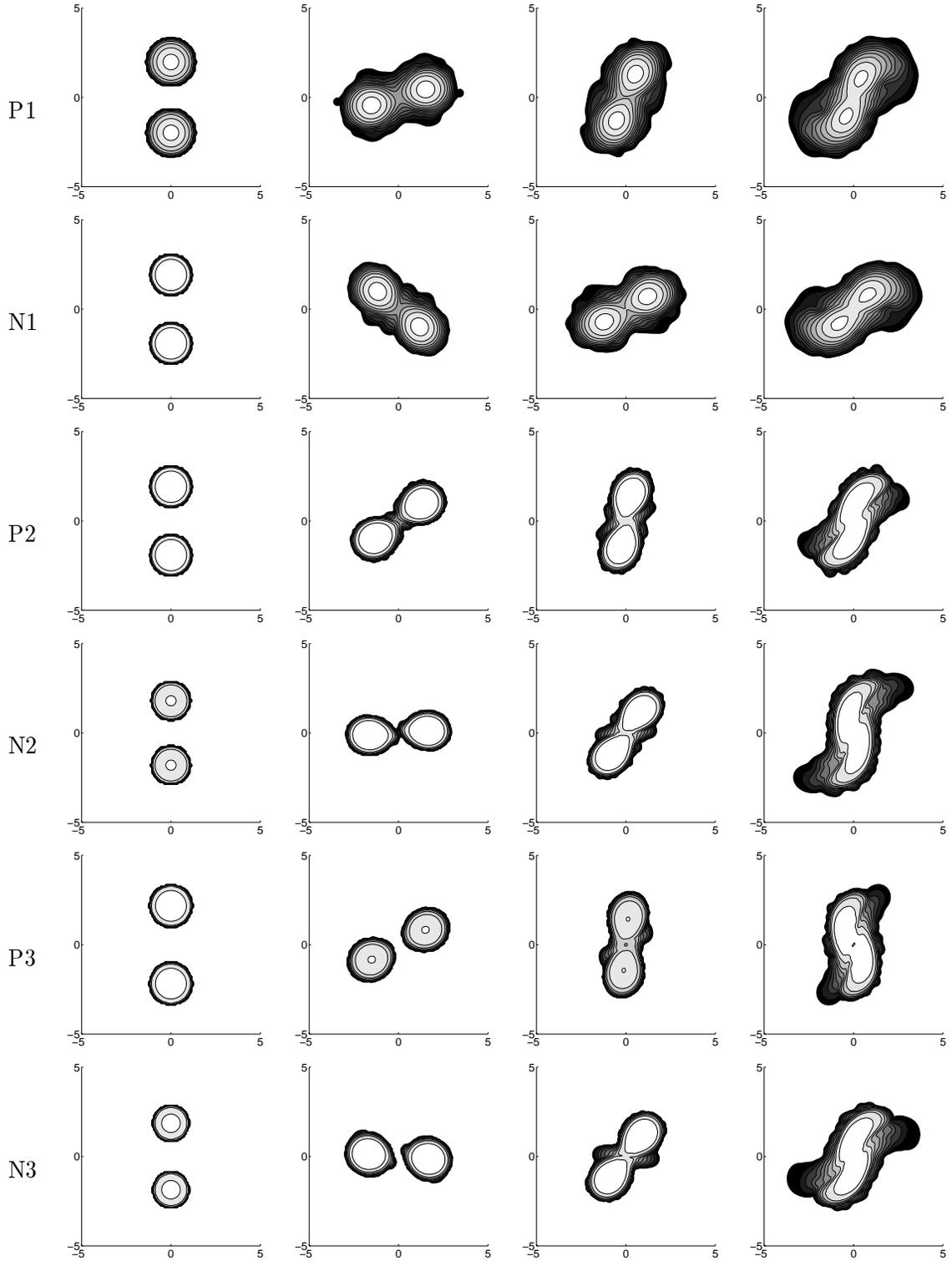}\\
    \caption{
      Coordinate rest mass density contours for the runs. Time
      increases to the right. The length scale is in $R_*$. The
      contours are taken at the initial time step, $t=-40$, $t=-25$
      (beginning of dynamical instability) and $t=-15$ ($a=2R_*$). The
      rest mass density is in units of $M/R_*^3$ and the contours are
      logarithmic with a spacing of $2.3$ starting at 1}
    \label{fig:dens1}
  \end{center}
\end{figure}

\begin{figure}
  \begin{center}
    \includegraphics{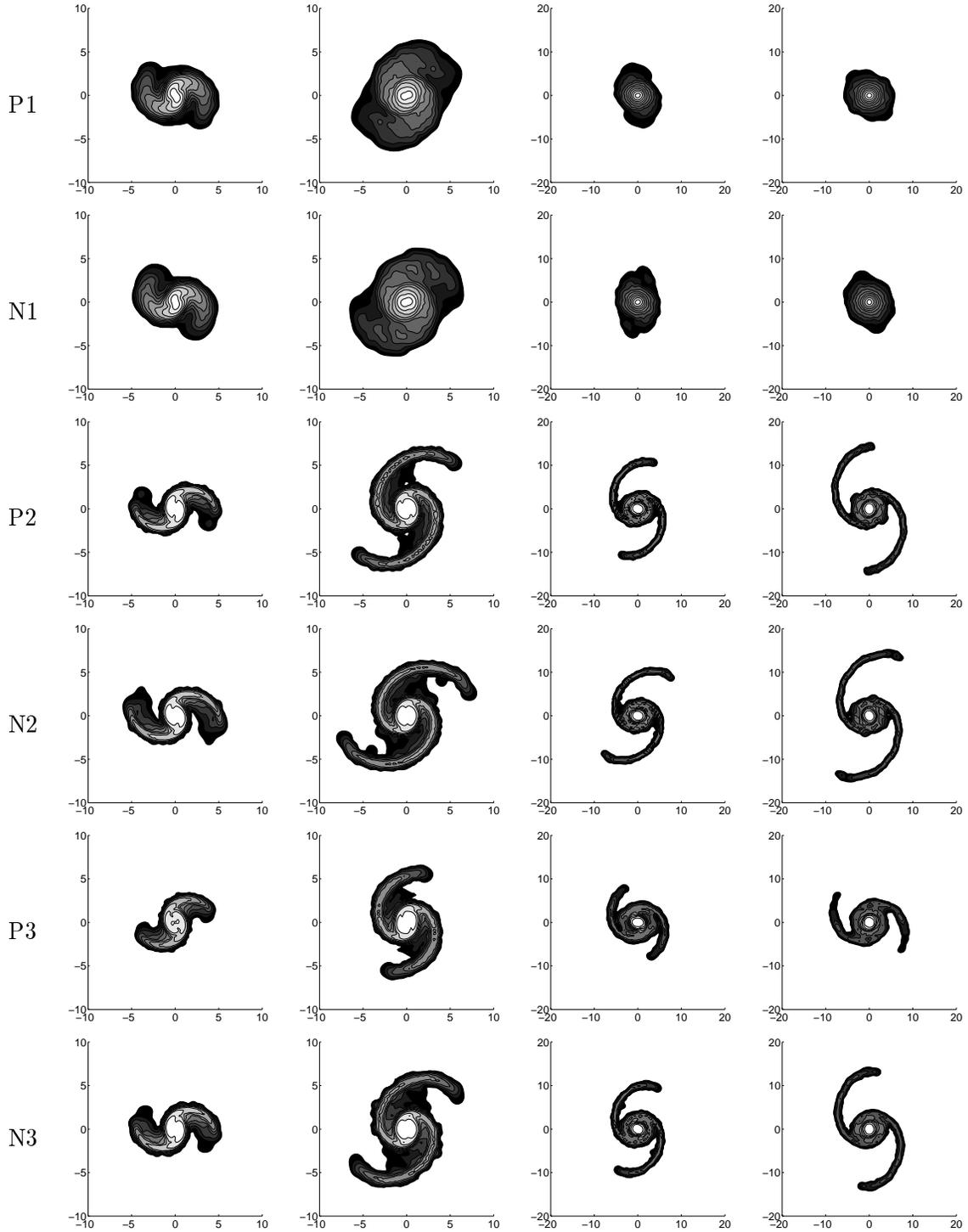}\\
    \caption{
      Coordinate rest mass density contours for the runs. Time
      increases to the right. The length scale is in $R_*$. The
      contours are taken at $t=-7$ (peak in Fig \ref{fig:mrs}c), $t=0$
      (minimal separation), $t=10$ and $t=20$.  The rest mass density
      is in units of $M/R_*^3$ and the contours are logarithmic with a
      spacing of $2.3$ starting at 1}
    \label{fig:dens2}
  \end{center}
\end{figure}

\begin{figure}
  \begin{center}
    \includegraphics{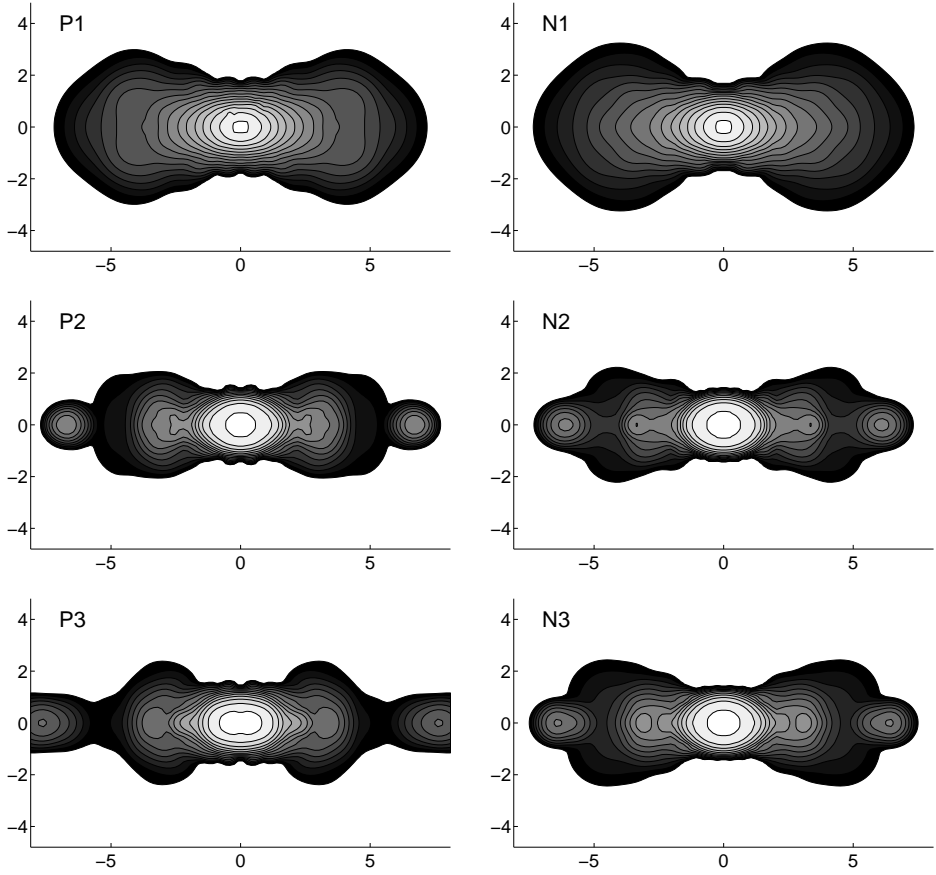}
    \caption{
      Coordinate rest mass density ($\rho_*$) in units of $(M/R_*^3)$
      for the final configuration ($t=20$) on the $x-z$ plane. The
      contours are logarithmic with a spacing of $2.1$
      starting at 1. The length scale is in $R_*$ }
    \label{fig:zcut}
  \end{center}
\end{figure}

We now turn to look at the morphological differences between the runs.
In Figures~\ref{fig:dens1} and ~\ref{fig:dens2} we show the contours
of coordinate rest mass density on the orbital ($x-y$) plane at
various times. The difference due to the different EoSs is the most
striking. The P1 and N1 runs lead to a final configuration with almost
non-existent spiral arms, which are very prominent in the P2, N2, P3,
and N3 runs.  There is almost no difference between the N1 and P1
results, while we see that N2 and N3 result in longer spiral arms than
P2 and P3 respectively. This is most prominent in the difference
between N3 and P3. The central object at $t=20$ is approximately the
same size in all the runs and is axisymmetric. For a closer look at
this central object we show contours of the rest mass density at t=20
in the $x-z$ plane (Fig. \ref{fig:zcut}). In all the runs we see a
central core, with a density above $0.1M/R_*^3$ and which has an
equatorial radius of approximately $1.5R_*$. The polar radius is
smaller in the N1 and P1 runs than in the P2, N2, P3, and N3 runs.
Surrounding this core is a halo with a radius of about $5R_*$.  The
halo is extended vertically in the N1 and P1 runs up to a distance of
$3R_*$, while in the P2, N2, N3, and P3 runs it has a height of only
$2R_*$. The P3 run resulted in the most compact object, as could be
expected.  In all runs, there is a funnel, a zone of low density,
around the axis of rotation.

Finally, we calculate how much mass can escape the system. We do this
by counting all the particles in the final configuration which are at
a distance greater than $6R_*$ from the origin and which have positive
total {\em Newtonian} energy. The total Newtonian energy of a particle
is $E_{\rm Newt} = \frac{1}{2}mv^2+m\varepsilon - mU_*$ with
$\varepsilon$ the specific internal energy and $U_*$ (Eq. \ref{eq:Us})
the Newtonian gravitational potential. Particles which are closer than
$6R_*$ will surely interact with other particles before exiting the
system and so their energy won't be conserved. For particles further
than this, the gravitational potential $U_*$ is of the order of 0.01
and the velocity is of the order of 0.1c which ensures that the
Newtonian energy will be conserved. Only in run N2, P2 and N3 have any
particles escaped, as shown in Table \ref{tab:parms}. The N2 has the
highest escaped mass followed by N3. For the other runs we are only
able to give upper bounds on the escaped mass by using the mass of the
lightest particles in the run. It's these minimal mass particles which
escape since they are located at the surface of the stars in the
initial conditions. We note that escaped mass is also given in
\cite{stephan}. However a comparison between our results is impossible
since we use different initial conditions and EoSs.

\section{Summary\label{sec:conc}}

We have introduced here a PN adaptation of the Tree+SPH algorithm.
This adaptation is made possible by the BDS formalism which recasts
the (0+1+2.5)PN equations of gravity and hydrodynamics in a form
resembling the Newtonian equations. We have tested various aspects of
our code by comparing the numerical results with known analytical
solutions of relativistic problems. The (0+1)PN hydrodynamic part had
been tested using a relativistic shock tube; the 2.5PN gravitational
radiation reaction terms were tested using two point masses in orbit
and the 1PN gravitation+hydrodynamic terms were tested using a
spherical static OV polytrope. The code passes all these tests with
the expected accuracy.

Using this code we have investigated the 1PN effects on BNS
coalescence. We compare runs with identical initial conditions but
different physics.  In the N runs, we use only the (0+2.5)PN terms, in
the P runs we include also the 1PN terms. In both runs we keep the
2.5PN gravitational radiation terms. These terms lead to a slow
decreases of the orbital separation until a critical separations is
reached. At this point a dynamical instability sets in and the stars
merge within one orbital period.  We use polytropes with a mass of
less than $M_\odot$ and a radius of about 30Km for the runs. These are
not typical NS parameters but the (0+1+2.5)PN approximation in the BDS
formalism is valid only when the 1PN terms are small compared to the
0PN terms and this set an upper limit to the compactness of the stars.
Therefore, Our results do not describe typical BNS coalescence, but
rather the effect of the 1PN terms on this process.

Our results show that when going up to higher masses and thus to more
relativistic conditions, there appears a prominent peak in the maximum
rest mass density just before the cores merge in the P3 run. This peak
is absent in the N3 run. This peak in rest mass density could mean a
that the probability of the coalesced object collapsing into a black
hole is larger than that estimated by Newtonian codes.  Also, we see
that the energy emitted in gravitational waves is almost twice as
large in the P3 run as compared with the N3 run. This difference is
also seen in the profile of the gravitational wave luminosity of the
system.  This result supports the suggested idea to use the
gravitational wave signal as a probe on the details of the merger
process. Our simulations show that the absence of a prominent second
peak in the luminosity indicates a soft EoS.

\appendix

\section{BDS Equations}
\label{sec:eqns}

We use the Lagrangian formulation of the BDS equations \citep{BDS}.
First we define some auxiliary quantities
\begin{eqnarray}
  \alpha &=& 2U_* -\gamma_*\left(\smfr{1}{2}\wsq + 3U_*\right),\label{eq:alpha}\\
  \beta  &=& \smfr{1}{2}\wsq + 3U_* + \varepsilon_* + \frac{ p_*}{\rho_*},\label{eq:beta}\\
  \delta &=& \smfr{3}{2}\wsq +  \varepsilon_*+3\frac{p_*}{\rho_*} - U_*,\label{eq:delta}\\
  A_i    &=& 4U_i + \smfr{1}{2}C_i - \smfr{1}{2}x^i\partial_sU_s,\\
  U_5    &=& \smfr{3}{2}\left( R-\Qd{ij}x^i\partial_jU_*\right),\\
  P_{ij} &=& 2\int d^3x\, \rho_*\left(
    3w_i\partial_jU_* -2\frac{w_i}{\rho_*}\partial_jp_* + 
    x^iw_s\partial_{sj}U_* - x^i\partial_{sj}U_s\right),\\
  \Qd{ij}&=& \smfr{1}{2}\left(P_{ij}+P_{ji}\right) - 
  \smfr{1}{3}\delta_{ij}P_{ss}\label{eq:Q},
\end{eqnarray}
where $p_*=p(e_*,\rho_*)$ is the pressure, $\gamma_*=\partial\log
p_*/\partial\log \rho_*$ is the adiabatic index and $\wsq=w_iw_i$. Using
these quantities we solve the following Poisson equations
\begin{eqnarray}
  \triangle U_* &=& -4\pi G\rho_* \label{eq:Us},\\
  \triangle U_i &=& -4\pi G\rho_*w_i, \\
  \triangle C_i &=& -4\pi Gx^i\partial_s\left(\rho_*w_s\right), \\
  \triangle U_2 &=& -4\pi G\rho_*\delta, \\
  \triangle R   &=& -4\pi G\Qd{ij}x^i\partial_j\rho_*.\label{eq:R}
\end{eqnarray}
The forces and the velocity are defined next
\begin{eqnarray}
  v^i      &=& w_i-\frac{1}{c^2}\beta w_i + \frac{1}{c^2}A_i + 
  \frac{4}{5}\frac{G}{c^5}w_s\Qd{is},\label{eq:vel}\\
  \Fr{press}_i &=& -\frac{1}{\rho_*}\partial_i
  \left[\left(1+\frac{\alpha}{c^2}\right)p_*\right],\\
  \Fr{1PN}_i &=& \left(1+\frac{\delta}{c^2}\right)\partial_iU_* +
  \frac{1}{c^2}\partial_iU_2-\frac{1}{c^2}w_s\partial_iA_s,\\
  \Fr{reac}_i &=& \frac{G}{c^5}\partial_iU_5.
\end{eqnarray}
We are now in position to advance the system in time using the
evolution equations
\begin{eqnarray}
  \dot{\rho_*} &=& -\rho_*\partial_iv^i \label{eq:cont},\\
  \dot{\varepsilon_*}   &=& -\frac{p_*}{\rho_*}\partial_iv^i, \label{eq:ener}\\
  \dot{x^i} &=& v^i,\\
  \dot{w_i} &=& \Fr{press}_i + \Fr{1PN}_i + \Fr{reac}_i \label{eq:moment},
\end{eqnarray}
where the dot represents the Lagrangian time derivative
$\dot{a}\equiv\partial_t a + v^i\partial_ia$.

\section{Adapting the Barnes-Hut tree to solve general Poisson equations}
\label{sec:ap1}
The Barnes-Hut tree is a method for calculating the gravitational
force between $N$ particles in $N\log N$ operations. The force on a
given particle is calculated by summing the forces from individual
particles if they are close, and by using a multipole approximation
for far away clusters of particles. A cluster is considered far away
if $d/l<\theta$ where $d$ is the clusters size, $l$ is it's distance
and $\theta$ is an external parameter governing the degree of
approximation of the method.  The tree itself is a data structure
specially adapted for efficient clustering of particles and
calculations of the multipole moments. In practice the multipole
approximation is accurate enough to be stopped at the quadrupole term.

Although it was originally invented for the calculation of
gravitational forces, the Barnes-Hut tree can solve any compact
supported Poisson equation with little modification. Given the
multipole moments of the source of the equation, the potential and
its derivatives can be calculated in an analogous way to the
calculation of the gravitational potential and force in the following
way:

Given a compact source $\eta$ and the masses and densities of each
particle, we will solve the Poisson equation
\begin{equation}
  \triangle U = -4\pi\eta.
\end{equation}
The first three multipole moments of a cluster of particles are
\begin{eqnarray}
  M^\eta   &=& \sum_p \frac{m_p}{\rho_p}\eta_p,\\
  D^\eta_i &=& \sum_p \frac{m_p}{\rho_p}\eta_p x^i_p,\\
  Q^\eta_{ij} &=& \sum_p \frac{m_p}{\rho_p}\eta_p
  \left(x^i_px^j_p-\smfr{1}{3}\delta_{ij}x^n_px^n_p\right),
\end{eqnarray}
where the index $p$ runs over the cluster particles, $m_p$ and
$\rho_p$ are the particle mass and density (so that $m_p/\rho_p\approx
V_p$ the particle volume) $\eta_p$ is the value of $\eta$ on the
particle and $x^i_p$ is the $i$-th component of the particle
coordinate.

Using these moments the cluster's potential and it's derivatives can
be calculated as follows:
\begin{eqnarray}
  U^\eta & = & \frac{M^\eta}{r} - \frac{D^\eta_ix_i}{r^3} + 
  \frac{1}{2}\frac{x_ix_jQ^\eta_{ij}}{r^5} \\
  \partial_kU^\eta 
  &=& -\frac{M^\eta}{r^3}x_k \nonumber\\
  &&  -\frac{1}{r^3}D^\eta_k + 3\frac{D^\eta_ix_i}{r^5}x_k \nonumber\\
  &&  +\frac{x_jQ^\eta_{jk}}{r^5} - 
  \frac{5}{2}\frac{x_ix_jQ^\eta_{ij}}{r^7}x_k \nonumber\\
  \partial_{lk}U^\eta 
  &=& -\frac{M^\eta}{r^3}\delta_{lk} + 3\frac{M^\eta}{r^5}x_kx_l \nonumber\\
  &&  +3\frac{1}{r^5}\left( D^\eta_kx_l + D^\eta_lx_k -
    D^\eta_ix_i\delta_{lk}\right)
  + 15\frac{D^\eta_ix_i}{r^7}x_kx_l \nonumber\\
  && + \frac{Q^\eta_{kl}}{r^5} - 10\frac{x_ix_lQ^\eta_{ik}}{r^7}
    -\frac{5}{2}\frac{x_ix_jQ^\eta_{ij}}{r^7}\delta_{kl} 
    +\frac{35}{2} \frac{x_ix_jQ^\eta_{ij}}{r^9}h_kx_l
\end{eqnarray}
where $r$ is the distance to the cluster, $\delta_{ij}$ is the
Kronecker delta and a double index is summed over.



\end{document}